\documentclass[useAMS,twocolumn,usenatbib]{mnras}
\pdfoutput=1
\usepackage{times,psfig,epsfig}
\usepackage{rotating, graphicx}
\usepackage{color}
\usepackage{amssymb, amsmath}
\usepackage[normalem]{ulem}
\usepackage[titletoc]{appendix}
\usepackage{adjustbox}

\def\aap{A\&A}

\def\apjl{ApJL}

\def\apjs{ApJS}

\newcommand{\bdm}{\begin{displaymath}}
\newcommand{\edm}{\end{displaymath}}
\newcommand{\beq}{\begin{equation}}
\newcommand{\eeq}{\end{equation}}
\newcommand{\beqnarr}{\begin{eqnarray}}
\newcommand{\eeqnarr}{\end{eqnarray}}
\newcommand{\bit}{\begin{itemize}}
\newcommand{\eit}{\end{itemize}}
\newcommand{\ben}{\begin{enumerate}}
\newcommand{\een}{\end{enumerate}}
\newcommand{\bfi}{\begin{figure}[htb]}
\newcommand{\bpfi}{\begin{figure}[p]}
\newcommand{\barr}{\begin{array}}
\newcommand{\earr}{\end{array}}
\newcommand{\bec}{\begin{center}}
\newcommand{\eec}{\end{center}}
\newcommand{\bs}{\begin{sideways}}
\newcommand{\es}{\end{sideways}}

 \newcommand{\mincir}{\raise
  -2.truept\hbox{\rlap{\hbox{$\sim$}}\raise5.truept \hbox{$<$}\ }}
\newcommand{\magcir}{\raise
  -2.truept\hbox{\rlap{\hbox{$\sim$}}\raise5.truept \hbox{$>$}\ }}
\newcommand{\siml}{\raise
  -2.truept\hbox{\rlap{\hbox{$\sim$}}\raise5.truept \hbox{$<$}\ }}
\newcommand{\simg}{\raise
  -2.truept\hbox{\rlap{\hbox{$\sim$}}\raise5.truept \hbox{$>$}\ }}

\title[Mass-Metallicity Relation]{Mass-Metallicity Relation from Cosmological Hydrodynamical Simulations and X-ray Observations of Galaxy Groups and Clusters}
\author[N. Truong et al]{
    N.~Truong$^1$\thanks{truongnhut@caesar.elte.hu},
 E.~Rasia$^{2,3}$,
V.~Biffi$^{2,4}$,
F. Mernier$^{1,8,9}$,
N. Werner$^{1,11,12}$,
M.~Gaspari$^{5}$\thanks{Einstein and Spitzer Fellow},
\newauthor
S.~Borgani$^{2,4,6}$,
S.~Planelles$^{7}$,
D.~Fabjan$^{10,2}$,
and G.~Murante$^{2}$
 \\~\\
\footnotesize
$^1$ MTA-E\"otv\"os University Lend\"ulet Hot Universe Research Group, P\'azm\'any P\'eter s\'et\'any 1/A, Budapest, 1117, Hungary \\
$^2$ INAF, Osservatorio Astronomico di Trieste, via Tiepolo 11, I-34131, Trieste, Italy \\
$^3$ Department of Physics, University of Michigan, 450 Church St., Ann Arbor, MI  48109 \\
$^4$ Dipartimento di Fisica dell' Universit\`a di Trieste, Sezione di Astronomia, via Tiepolo 11, I-34131 Trieste, Italy \\
$^{5}$ Department of Astrophysical Sciences, Princeton University, 4 Ivy Ln, Princeton, NJ 08544-1001, USA\\
$^{6}$ INFN, Instituto Nazionale di Fisica Nucleare, Trieste, Italy \\
$^{7}$ Departamento de Astronom\'ia y Astrof\'isica, Universidad de Valencia, c/Dr. Moliner, 50, 46100 Burjassot, Valencia, Spain \\
$^{8}$ Institute of Physics,E\"otv\"os University, P\'azm\'any P. s. 1/A, Budapest, 1117, Hungary \\
$^{9}$ SRON Netherlands Institute for Space Research, Sorbonnelaan 2, 3584 CA Utrecht, The Netherlands \\
$^{10}$ Faculty of Mathematics and Physics, University of Ljubljana, Jadranska 19, 1000 Ljubljana, Slovenia \\
$^{11}$ Department of Theoretical Physics and Astrophysics, Faculty of Science, Masaryk University, Kotl\'a\v{r}sk\'a 2, Brno, 611 37, Czech Republic \\
$^{12}$ School of Science, Hiroshima University, 1-3-1 Kagamiyama, Higashi-Hiroshima 739-8526, Japan \\
}

\begin{document}
\maketitle

\begin{abstract}
Recent X-ray observations of galaxy clusters show that the distribution of intra-cluster medium (ICM) metallicity is remarkably uniform in space and time. 
In this paper, we analyse a large sample of simulated objects, from poor groups to rich clusters, to study the dependence of the metallicity and related quantities on the mass of the systems.
The simulations are performed with an improved version of the Smoothed-Particle-Hydrodynamics \texttt{GADGET-3} code and consider various 
astrophysical processes including radiative cooling, metal enrichment and feedback from stars and active galactic nuclei (AGN). The scaling between the metallicity and the temperature obtained in the simulations agrees well in trend and evolution with the observational results obtained from two data samples
characterised by a wide range of masses and a large redshift coverage.  We find that the iron abundance in the cluster 
core ($r<0.1R_{500}$) does not correlate with the temperature nor presents a significant evolution. The scale invariance is confirmed when the metallicity
is related directly to the total mass. The slope of the best-fitting relations is shallow ($\beta\sim-0.1$) in the innermost regions ($r<0.5R_{500}$) and consistent with zero outside. 
We investigate the impact of the AGN feedback and find that it plays a key role in producing a constant value of the outskirts metallicity from groups to clusters.
This finding additionally supports the picture of early enrichment.

\end{abstract}

\section{Introduction}\label{sec:intro} 

Being the largest gravitationally-bound objects in the Universe, galaxy clusters can in first approximation be considered as closed boxes which contain a fair representation of the cosmic baryon content (\citealt{white.etal.1993}, \citealt{frenk.etal.1999}). Therefore they are ideal laboratories to study the cosmic cycle of the baryonic matter in its various phases: hot intra-cluster medium (ICM), cold gas, and stellar component. The first component, the ICM, emits predominantly in the X-ray band due to its high temperature ($T\sim10^{7}-10^{8}$ K), making X-ray observations a key tool to study thermodynamics of the ICM (see \citealt{bohringer.werner.2010} for a review). Furthermore, emission-line features of the X-ray observed spectra also reveal a wealth of information about the chemical composition of the intra-cluster gas, thereby offering a unique window to study the ICM metal enrichment (\citealt{werner.etal.2008,mernier18_REVIEW}).

The ICM metal enrichment involves numerous astrophysical processes (see, e.g., \citealt{borgani.etal.2008,biffi18_REVIEW}). Metals, elements that are heavier than H and He, are created via stellar nucleosynthesis and released by means of stellar mass loss of low- and intermediate-mass stars or of supernova (SN) explosions. Different types of supernovae produce different elements. The core-collapse or Type II supernovae (SNII) produce mainly light elements, e.g. O, Ne, Mg, or Si, while Type-Ia supernovae (SNIa) are source of heavy metals: Fe and Ni. The intermediate-mass elements are produced by both types of supernovae, and lighter elements (C, N) originate from low- and intermediate-mass stars during their Asymptotic Giant Branch (AGB) phase. The produced metals subsequently enrich the surrounding astrophysical environment
thanks to multi-scale mixing processes, such as galactic winds, active galactic nuclei (AGN) feedback, ram-pressure stripping, and mergers (see \citealt{schindler.diaferio.2008} for a review). In particular, X-ray observations show evidences that AGN outflows can eject metals up to several 100 kpc even in massive galaxy clusters and groups (e.g., \citealt{kirkpatrick.etal.2011,ettori.etal.2013}). This phenomenon is reproduced in high-resolution hydrodynamical simulations as well \cite[e.g.,][]{gaspari.etal.2013}.

Despite the astrophysically complex nature of the ICM metal enrichment, recent X-ray observations have shown that the distribution of metals in the ICM is remarkably homogeneous in space (\citealt{werner.etal.2013,urban.etal.2017}) and in time (\citealt{ettori.etal.2015,McDonald.etal.2016}). \cite{werner.etal.2013} and \cite{urban.etal.2017} show that the iron profile in the ICM of Perseus and another ten clusters observed by Suzaku is flat at radii $r>0.25R_{200}$\footnote{The mass enclosed within a sphere that has averaged density equal to $200$ times the critical density of the Universe ($\rho_{\rm{crit}}$). In general, the mass $M_\Delta$ is related to the radius $R_\Delta$ by the equation: $M_\Delta=\frac{4\pi}{3}\Delta\rho_{\rm{crit}}R_\Delta^3$. We also use $R_{500}$ and $R_{2500}$ in this work.} with an iron abundance level of $\sim0.3\ Z_{Fe,\odot}$. In another work, \cite{McDonald.etal.2016}, X-ray observations of $153$ clusters observed by {\it Chandra}, {\it XMM-Newton}, and {\it Suzaku} telescopes, reveal no evidence for strong redshift evolution of the global ICM metallicity ($r<R_{500}$) (see also \citealt{ettori.etal.2015}). Even more interestingly, a recent observation of the Perseus cluster with {\it Hitomi} (\citealt{hitomi.2017}) shows that the chemical composition of its ICM is consistent with the solar chemical composition in terms of abundance ratios.

Numerical modelling also supports the uniform picture of ICM metal enrichment. In recent studies based on simulations, \cite{biffi.etal.2017, biffi.etal.2018} and \cite{vogelsberger.etal.2018} also find a flat metallicity profile in cluster outskirts that is constant over time. The uniformity of ICM metallicity in space and time supports the picture of early metal enrichment. In this scenario, early-time pristine gas in galaxies was enriched at high redshift ($z\sim5-6$) and then spread widely mainly by feedback processes. The gas was subsequently accreted into massive halos and heated by gravitational compression. Since the gas was already metal-rich at the time of accretion, the ICM metallicity profile appears to be flat at large cluster-centric distances.

In addition to a flat and constant metal distribution, the early enrichment scenario can be tested by investigating the mass dependence of the metallicity. For this reason, there
is recently interest from both theoretical (\citealt{dolag.etal.2017,yates.etal.2017, barnes.etal.2017,vogelsberger.etal.2018}) and observational (\citealt{renzini.andreon.2004,mantz.etal.2017, yates.etal.2017}) sides in investigating how the ICM metallicity varies with cluster scale. Up to date, the most complete observed sample, for the study of averaged ICM metallicity in clusters, has been carried out by \cite{mantz.etal.2017}. This work shows that iron abundance in the ICM anti-correlates with ICM temperature, however, the study is limited to a massive sample only ($T>5$ keV). An earlier work from \cite{yates.etal.2017}, based on a compilation of various observational datasets from the literature and on a sample obtained from semi-analytical models, reports a similar anti-correlation between observed iron abundance and temperature in high-temperature systems, while in low-temperature systems ($T<1.7$ keV) the compiled sample exhibits a drop in the iron abundance (see also, \citealt{sun.etal.2012,mernier.etal.2016}).  This drop is not present in recent theoretical studies (e.g., \citealt{yates.etal.2017}).

In this paper we carry out a thorough study of how ICM metallicity varies
(1) from group to cluster scales,
(2) in various radial ranges
and (3) in time, by employing cosmological hydrodynamical simulations. 

The simulated sample that we analysed is an extended dataset with respect to the one presented in \cite{biffi.etal.2017,biffi.etal.2018}.
Those simulated clusters have also been shown to reproduce various realistic thermodynamical properties of the ICM (\citealt{rasia.etal.2015, Villaescusa-Navarro.etal.2016, biffi.etal.2016, planelles.etal.2017, truong.etal.2018}).

In the first part of this study, we present a comparison between simulations and observations on the dependence of the ICM metallicity on the gas temperature, in which specific attention is payed to potential biases that might affect the observational results.
In particular, the observational datasets that we use include the CHEERS sample observed by the {\it XMM-Newton} telescope (\citealt{de_plaa.etal.2017}), in which the X-ray spectra are analysed with the last up-dated model of ICM emission for metallicity estimation. The detailed description of the CHEERS data as well as its X-ray analysis is provided in a companion paper (\citealt{mernier.etal.2018}). In addition, we also compare our simulation results against observational data by \cite{mantz.etal.2017}.

In the second part, we theoretically investigate how the metallicity varies as a function of cluster mass as well as how this relation evolves over redshift. In order to enhance the role played by AGN feedback on the early enrichment, we study how the iron mass, hydrogen mass and stellar fraction relate with the total mass in simulations with and without AGN feedback.

The paper is organized in the following way. We describe the main features of the numerical simulation and the method of analysis in Section 2. In Section 3, we briefly introduce the observed datasets: the CHEERS (\citealt{de_plaa.etal.2017}), and \cite{mantz.etal.2017} samples. This is followed by a detailed comparison between simulations and observations. In Section 4, we present a theoretical investigation of the iron and oxygen abundance-mass relations as well as their evolution over time. The effects of AGN feedback on the mass-metallicity relation, and related quantities, are also discussed in this Section. We devote Section 5 to a detailed discussion of the systematics of simulation results and to a comparison with other theoretical studies. Finally, we summarise the main results and conclude in Section 6.

Throughout this study, we adopt the solar metallicity values provided by \cite{asplund.etal.2009} for both the theoretical and the observational analysis and we consider $h\equiv H_0/(100\ \rm{km}\ \rm{s}^{-1}\ \rm{Mpc}^{-1})=0.72$, where $H_0$ is the Hubble constant.

\label{sec:cmrr}

\vspace{0.3cm}

 \section{Analysis of Simulations}
\label{Simulation}
\subsection{The Simulated Clusters}
The simulations 
employed in this study have been described 
in recent works (\citealt{rasia.etal.2015, Villaescusa-Navarro.etal.2016, biffi.etal.2016,planelles.etal.2017, biffi.etal.2017, biffi.etal.2018, truong.etal.2018}),
where we show that they reasonably reproduce chemo- and thermo-dynamical cluster properties in comparison to observational data.

In this Section we summarise only the main features and refer the reader to the aforementioned references for more details, and specifically to \cite{biffi.etal.2017} for the description of how chemical enrichment is included in the simulations.

The simulated galaxy clusters are obtained from $29$ zoomed-in Lagrangian regions re-simulated with an upgraded version of the \texttt{GADGET-3} code (\citealt{springel.2005}). Starting from an initial Dark-Matter (DM) only simulation with a volume of $(1 h^{-1}\rm{Gpc})^3$, 24 regions around most massive clusters with mass $M_{200}>8\times10^{14}h^{-1}M_\odot$, and 5 regions  surrounding isolated groups with $M_{200}\sim[1-4]\times10^{14}h^{-1}M_\odot$ are selected and re-simulated at higher resolution and with added baryonic components. Each high-resolution region has a radius that extends at least 5 times the virial radius of the central object covering a box size of about $25-30 h^{-1}$ Mpc (see \citealt{bonafede.etal.2011} for more details on the initial conditions). The mass resolutions for the DM particles and the initial gas particles are $m_{DM}=8.47\times10^{8}h^{-1}M_\odot$, $m_{gas}=1.53\times10^{8}h^{-1}M_\odot$, respectively. The  simulation is performed with cosmological parameters consistent with results from the WMAP-7 (\citealt{komatsu.etal.2011}): $\Omega_m=0.24$, $\Omega_b=0.04$, $n_s=0.96$ for the primodial spectral index, $\sigma_8=0.8$ for the amplitude of the density fluctuations power spectrum, and $H_0=72\ \rm{km}\ \rm{s^{-1}}\ \rm{Mpc^{-1}}$ for the Hubble parameter. The Plummer-equivalent softening length is fixed equal to $\epsilon=3.75 h^{-1}\rm{kpc}$ for DM and gas particles, and $\epsilon=2h^{-1}\rm{kpc}$ for black hole and star particles. The DM softening length is fixed in comoving units for $z>2$ and in physical units at lower redshifts. For other types of particles, it is always given in comoving units.

Regarding the hydrodynamical scheme, we employ the improved Smoothed-Particle-Hydrodynamics (SPH) formulation described in \cite{beck.etal.2016}. Comparing to the standard \texttt{GADGET} code, the new SPH scheme incorporates a number of advanced features including: the choice of a higher-order Wendland $C^4$ kernel function, the implementation of a time-dependent artificial viscosity scheme, and artificial conduction. These advanced features improve the scheme's ability in treating contact discontinuities and gas-dynamical instabilities, thereby overcoming several limitations of standard SPH schemes.

For the study of metallicity of the intra-cluster gas, we adopt two different prescriptions for the ICM physics which have been used to produce two simulated samples from the same initial conditions. The run called CSF (cooling-star-formation) includes the following physical processes:

\begin{itemize}
\item Heating/cooling from Cosmic Microwave Background (CMB) and from a UV/X-ray time-dependent uniform ionising background included as in \cite{haardt.madau.2001}.

  \item Metallicity-dependent radiative gas cooling as in \cite{wiersma.etal.2009}, and star formation~\cite[][]{springel.hernquist.2003}.

\item Metal enrichment, as described in \cite{tornatore.etal.2007}, accounting for three different channels of enrichment, namely from SNII, SNIa, and AGB stars. We follow the production and evolution of 15 different chemical species: H, He, C, Ca, O, N, Ne, Mg, S, Si, Fe, Na, Al, Ar, Ni. We assume the stellar initial mass function (IMF) by \cite{chabrier.2003}, and the mass-dependent lifetimes of \cite{padovani.matteucci.1993}. The stellar yields used in the model are the set provided in \cite{thielemann.etal.2003} for SNIa stars and the one from \cite{karakas.2010} for AGB stars. For SNII, we use the metal-dependent yields taken from \cite{woosley.weaver.1995} and \cite{romano.etal.2010}. The heavy elements produced by stellar particles in the simulations, according to this model, are then distributed to the surrounding gas particles by smoothing them onto the SPH kernel, as it is done for the other thermodynamical quantities. Once gas particles are enriched, the diffusion of metals is essentially due to the motion of the gas particles themselves. This allows metals to be circulated on large cluster scales over time.

\item Thermal feedback from supernovae as originally prescribed by \cite{springel.hernquist.2003} with the value of the mass-loading parameter equal to 2. Kinetic feedback from the SN is also included, in the form of galactic winds with a velocity of $350\ \rm{km}\ \rm{s}^{-1}$.

\end{itemize}
\begin{figure}
\centering
{\includegraphics[width=0.5\textwidth]{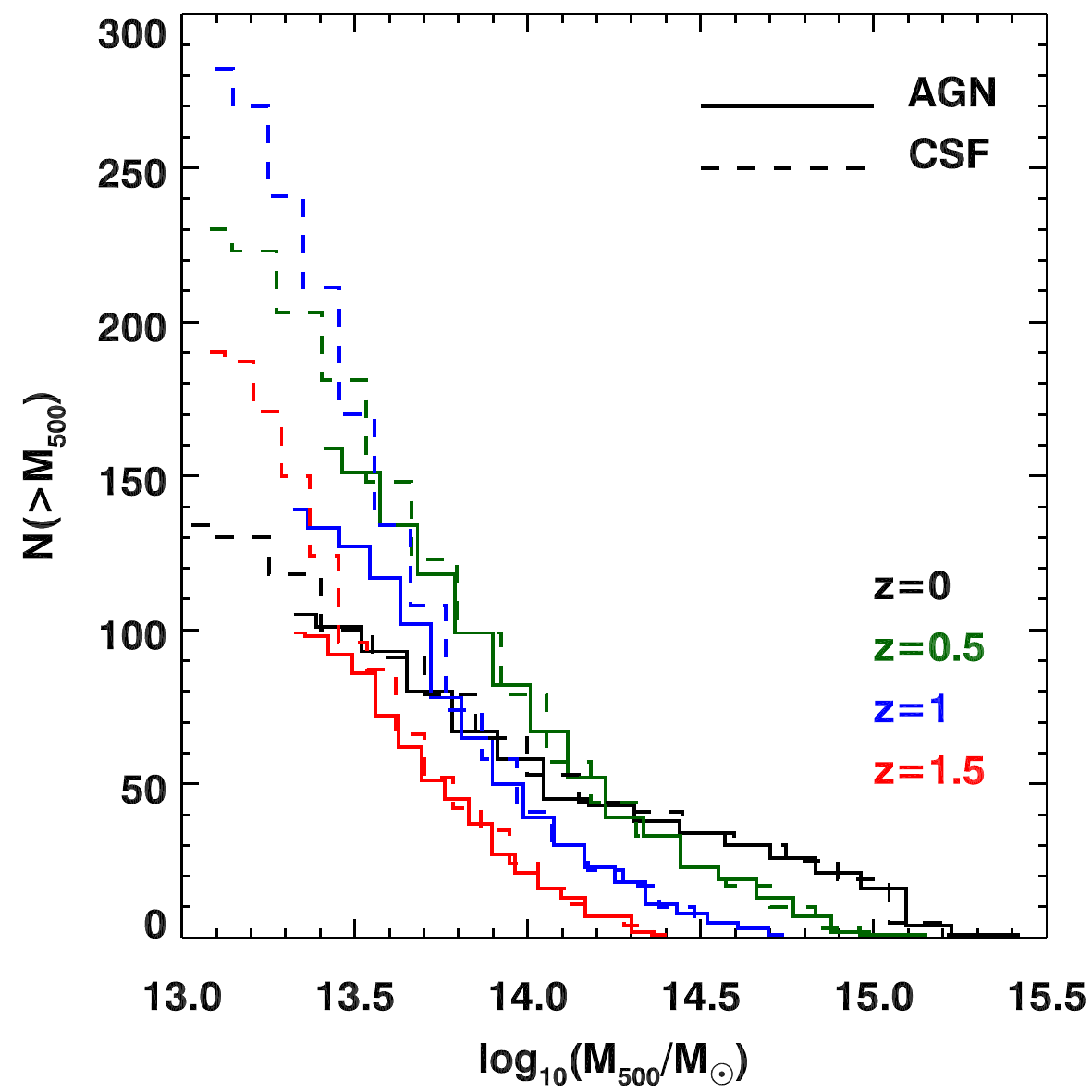}}
\caption{The cumulative distribution of $M_{500}$ in our AGN and CSF simulations at different redshifts for those objects with $M_{500}>10^{13}M_\odot$ and having at least 100 gas particles in the core ($r<0.1R_{500}$). The total number of selected objects in the AGN (CSF) simulation at each redshift is: 105 (134), 159 (230), 139 (282), and 99 (190) for $z=0$, $0.5$, $1$, and $1.5$, respectively.}
   \label{fig1}
\end{figure}
The other run is called AGN. It has all the features of the CSF run but additionally includes the treatment of gas accretion onto super-massive black holes (SMBH) powering AGN feedback, for which we employ the model by \cite{steinborn.etal.2015}
(an upgrade from the original model proposed by \citealt{springel.dimatteo.hernquist.2005}) that considers both hot and cold accretion (e.g., \citealt{gaspari.etal.2018}). In this updated model, we consider both radiative and mechanical feedback that are produced via gas accretion onto SMBH, and both are released into the surrounding environment in form of thermal energy. The radiated energy then couples with the surrounding gas with feedback efficiency $\epsilon_f=0.05$. In this model, we treat cold and hot gas accretion separately, in particular for the cold gas accretion, we boost the Bondi rate by a factor of $100$ thereby mimicking the effect of the cold accretion mode (see more detailed discussion in \citealt{gaspari.temi.brighenti.2017}). For this study, we select a mass-limited sample with $M_{500}>10^{13}M_\odot$ with at least $100$ gas particles in the core regions ($r<0.1R_{500}$)\footnote{The fraction of AGN (CSF) clusters with at least $100$ gas particles in the core with respect to the total sample are: $34 (39)$, $39(53)$, $34(64)$, and $36(67)$ per cent, at $z=0$, $0.5$, $1$, and $1.5$, respectively.}. In Fig.~\ref{fig1} we show the cumulative distribution of cluster total mass for the AGN and CSF simulations, in terms of $M_{500}$, at the four redshifts, $z=0, 0.5, 1$, and $1.5$. 

\subsection{Computing simulated ICM quantities}

In the following we briefly describe how quantities of interest are computed within the simulation snapshots. In computing X-ray quantities, we select only gas particles with a low fraction of cold gas ($<10\%$) to avoid the inclusion of star-forming particles.

\begin{itemize}

\item \emph{Projected quantities.}
  For the purpose of comparing simulation results with
  observational data, we employ projected quantities.  Namely,
  we compute the temperature and the metallicity within a cylindrical volume with the length of
  $2\times R_{vir}$, where $R_{vir}$ is the virial radius\footnote{For
    the cosmology used in our simulations, the virial radius, according to \cite{bryan.norman.1998}, 
    corresponds to $\Delta\approx93$ at $z=0$.}, and the area
  confined by two circular apertures.  In order to be
  consistent with X-ray observations, which are typically centred on the
  surface brightness peak, we centre the cylindrical regions on the centre of the gas mass
  calculated from the gas particles within $R_{2500}$ (which is typically $\sim 0.3 R_{500}$). We verify that the impact of using the gas centre of mass instead of the X-ray peak is negligible in the case of the metallicity-temperature scaling (see Appendix A2).

For the temperature, we adopt the spectroscopic-like
formula proposed by~\cite{mazzotta.etal.2004}:
\begin{equation}
T_{\rm sl}=\frac{\Sigma_i \rho_i m_i T_i^{0.25}}{\Sigma_i \rho_i m_i T_i^{-0.75}},
\end{equation}
where $m_i$, $\rho_i$, and $T_i$ are the mass, density, and
temperature of the $i^{th}$ gas element, respectively. For the
computation of spectroscopic-like temperature, we apply a temperature
cut $T_{i}>0.3$ keV to select particles that should emit in the X-ray band.

For the metallicity, we adopt the {\it emission-weighted}
estimate. In case of iron (and similarly for oxygen) this quantity is defined as:
\begin{equation}
Z_{\rm Fe,ew}=\frac{\Sigma_i n_{e,i}n_{H,i}\Lambda(T_i, Z_i)\times Z_{i,{\rm Fe}}}{\Sigma_i n_{e,i}n_{H,i}\Lambda(T_i, Z_i)},
\end{equation}
where $Z_i$ and $Z_{i,{\rm Fe}}$ are respectively the global and the iron metallicities of the gas particle, and $n_{e,i}$, $n_{H,i}$
are the electron and hydrogen number densities,
respectively. $\Lambda$ is the cooling function computed based on
particle temperature and metallicity by assuming the APEC model
(\citealt{smith.etal.2001}) in XSPEC\footnote{https://heasarc.gsfc.nasa.gov/docs/xanadu/xspec/} v.12.9.0 (\citealt{arnaud.1996}) for radiative emission and by
integrating over the $[0.01-100]$ keV energy band.

\item \emph{Theoretical estimates.}
  For the theoretical investigation of the simulated clusters, we
  use three-dimensional measurements, computed within spherical
 shells centred on the minimum of the system potential well.

  For the iron and oxygen abundances we
  employ the {\it mass-weighted} formula, defined as:
  \begin{equation}
    Z_{\rm X,mw}=\frac{\Sigma_i m_i\times Z_{i,{\rm X}}}{\Sigma_i m_i},
  \end{equation}
  where $m_i$ is the gas particle mass, and $\rm X$ represents one of the two elements.

  Similarly, to explore the intrinsic dependence of ICM metallicity on
  the cluster scale and related quantities, we also measure total,
  iron and hydrogen masses, as well as stellar fractions, evaluated
  within spherical three-dimensional regions.

\end{itemize}

\subsection{Fitting Method}

 We model the relations between metallicity, total mass or temperature, and redshift   
 with power-law functional forms. The specific function adopted will be specified, case by case, in the following sections.
  In order to characterise the slope and normalisation of the relations,   we always perform 
  a log-log linear regression fit.  For this task, we employ the IDL routines \texttt{linmix\_err.pro} 
  and \texttt{mlinmix\_err.pro} which adopt a Bayesian approach to linear regression, in which best-fit parameters are determined via a Monte Carlo Markov Chains method, as described 
  in \cite{kelly.2007}. This method enables us to treat the intrinsic scatter as a free parameter, like 
  the slope and normalisation, and to account for possible correlation between measurement errors and intrinsic scatter (see \citealt{kelly.2007} for detailed discussion). However, it is worth noticing that for simulation data, which do not have any associated statistical uncertainty, there are simpler methods that can be used for the fitting. For instance, we find that, by using the non-Bayesian fitting routine \texttt{robust\_linefit.pro} in IDL to fit the mass-metallicity relation ($Z_{Fe}-M_{500}$) in the central region ($r<0.1R_{500}$), the results are consistent with the Bayesian approach within $1\sigma$.
  \vspace{0.3cm}

%
\section{Comparison to observations}
In the first part of this study we investigate the iron abundance in
simulated and observed clusters as a function of the ICM temperature.
In fact, even though the mass is the optimal description of the system scale, this quantity is not directly
  observable and temperature measurements, which are easier to derive, are
  typically used in X-ray studies.  For the purpose of a
  more faithful comparison to observational data, we therefore employ
  projected spectroscopic-like temperature estimates of the ICM in
  simulated clusters as well.
  We note that the relation between temperature and mass for the
  clusters in our AGN simulations is in reasonable agreement with
  observational findings, both at low and intermediate redshifts \cite[][]{truong.etal.2018}.

\subsection{Observational Data Sets}
\label{Observation}

In the following we briefly describe the main
  features of the two observational datasets used for the comparison
  with simulation results, namely the CHEERS sample and the
 sample presented in \cite{mantz.etal.2017}.
Specifically, we will take advantage of the large temperature range
spanned by the former, a local cluster sample, in order to
investigate in detail the dependence of central metallicity on
core temperature in local clusters.
For the purpose of exploring the redshift evolution of the
metallicity-temperature relation, in different radial ranges,
we will use instead the latter, which contains a larger number of massive clusters.

  \begin{figure*}
\centering
{\includegraphics[width=0.7\textwidth]{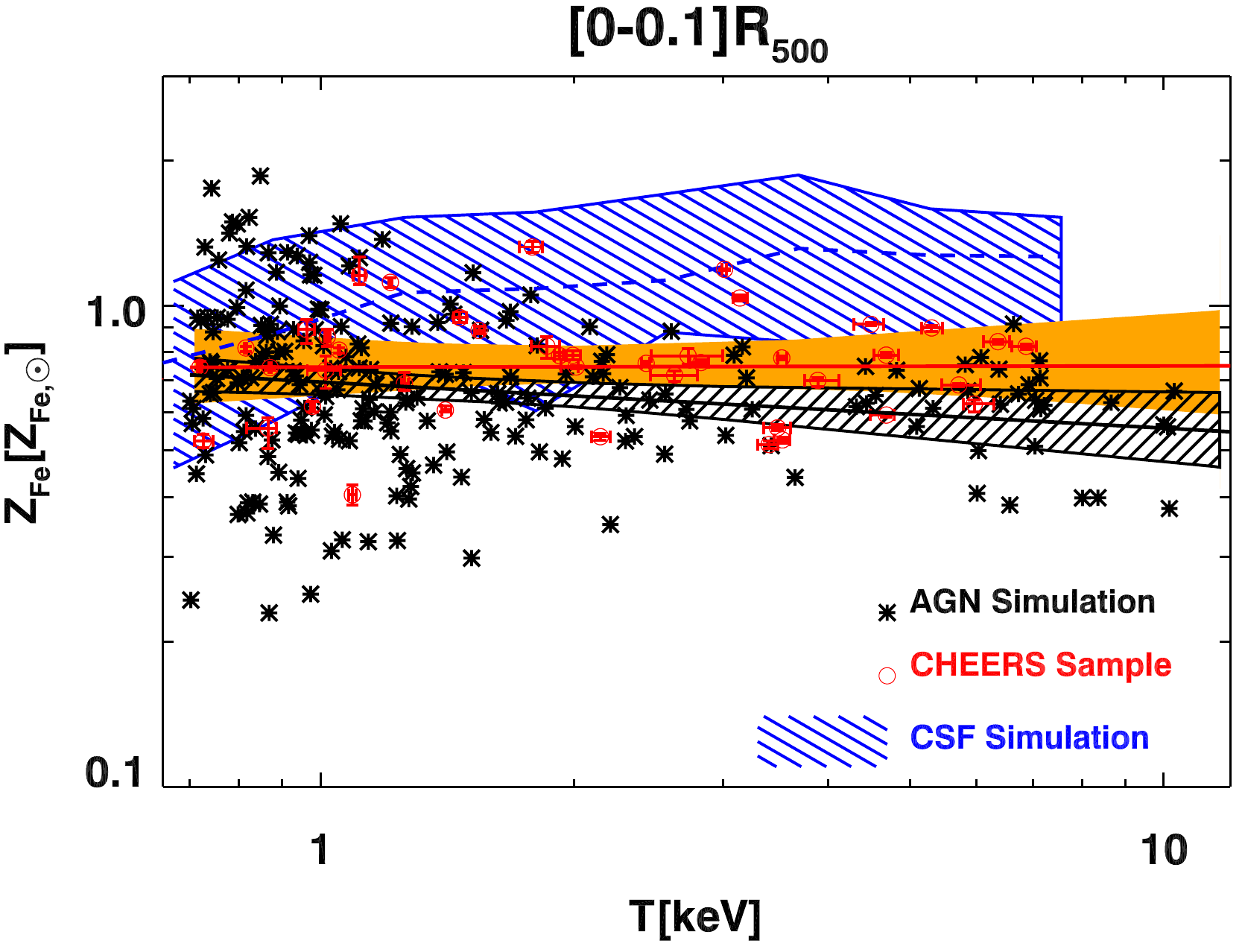}}
\caption{Simulated and observed $Z_{\rm Fe}-T$ relations in the cluster core ($r<0.1R_{500}$): in red circles are the observational results from the CHEERS sample, in black asterisks are the measurements from the AGN simulation, while the blue dashed line and the blue-shaded area represent the median relation and the $68.3\%$ confidence region, respectively, for the CSF simulation. For the AGN and CHEERS samples, we show the best-fit relations (black and red solid lines) as described by Eq.~(\ref{eq4}) and the associated $68.3\%$ confidence regions (black- and orange-shaded areas).} %
   \label{fig2}
\end{figure*}
\begin{itemize}
\item The CHEERS sample includes 43 nearby ($z<0.1$) cool core clusters, groups, and massive ellipticals observed by {\it XMM-Newton}. The iron abundance is constrained by the {\it XMM-Newton} EPIC (MOS 1, MOS 2, and pn) instruments, as they can access both the Fe-L and the Fe-K line complexes of the X-ray spectral window. We use the Fe measurements from \cite{mernier.etal.2018}, which are derived within $0.1R_{500}$ and obtained from fitting X-ray spectra using the up-to-date version of the spectral code used to model the ICM emission, namely SPEXACT\footnote{SPEX Atomic Code and Tables, as part of the fitting package SPEX (https://www.sron.nl/astrophysics-spex).}. A major update has been recently released (i.e., from SPEXACT v2 to SPEXACT v3, \citealt{de_plaa.etal.2017}) and, compared to the previous observational results, SPEXACT v3 revises the Fe abundance in groups significantly higher and makes them on average consistent with that in clusters.
In order to minimise the impact of the Fe-bias, all the spectra were fitted with three single-temperature components that mimic a Gaussian temperature distribution. The complete details and discussion on the data analysis methods and the effects of the latest spectral model improvements are presented in \cite{mernier.etal.2018}.
\item The sample by \cite{mantz.etal.2017} is the largest cluster sample available up-to-date. It consists of 245 massive systems ($T>5\,$keV) selected from X-ray and Sunyaev-Zel'dovich effect surveys, with X-ray observations obtained from {\it Chandra}. The sample encompasses a broad redshift range: $0<z<1.2$. Metallicity analysis is performed for three radial ranges: $[0-0.1]R_{500}$, $[0.1-0.5]R_{500}$, and $[0.5-1]R_{500}$. The spectrum extracted in each annular bin is fitted using XSPEC assuming a single-temperature component APEC model (ATOMDB version 2.0.2).
\end{itemize}

\label{Comparison to observations}

\subsection{The $Z_{\rm Fe}-T$ relation for groups and clusters}

In Fig.~\ref{fig2} we show the $Z_{\rm Fe}-T$ relation measured within $0.1R_{500}$ for our simulated samples in comparison to the CHEERS dataset.
Since the observational sample includes only nearby objects, we consider only $z=0$ simulated clusters. In addition, to be consistent with the data, we
select simulated systems that have temperature greater than $0.7$ keV at $z=0$. For the sake of clarity, we show the individual data points for the AGN (black asterisks) and CHEERS (red circles) samples, while for the CSF sample we present only the median $Z_{\rm Fe}-T$ relation and the associated $68.3\%$ confidence region (blue dashed line and shaded area).

To characterise the correlation between $Z_{\rm Fe}$ and $T$, we compute the Spearman's rank correlation coefficients ($r_s$) for simulated and observed samples. Additionally, we also quantify the dependence of iron abundance on temperature by fitting the simulated and the observed data with the following formula:
\begin{equation}
Z_{\rm Fe}=Z_{0T}\bigg(\frac{T}{1.7\ \rm{keV}}\bigg)^{\beta_T}, \label{eq4}
\end{equation}
which is characterised by two parameters: the normalisation $Z_{0T}$ and the slope $\beta_T$. We choose the pivot point for the temperature equal to $1.7$ keV, which is close to the mean values of the simulated and observed sample temperatures. The $Z_{\rm Fe}-T$ in the CSF run cannot be simply described by a single power law (see the discussion in the next paragraph), therefore we fit Eq.~(\ref{eq4}) separately to high-temperature ($T>1.7$ keV) and low-temperature ($T<1.7$ keV) subsamples. The correlation coefficients $r_s$ and the best-fitting parameters of Eq.~(\ref{eq4}) are reported in Table~\ref{tb1}.
 \begin{table*}
  \caption{\label{tb1}
  Best-fitting parameters of the $Z_{\rm Fe}-T$ relation in the clusters core ($[0-0.1]R_{500}$), as described by Eq.~(\ref{eq4}), for different AGN and CSF simulated samples shown along with the observational CHEERS sample. The Spearman's rank correlation coefficients ($r_s$) and the null-hypothesis probability (in parentheses) are also reported.}
 \begin{center}
 \begin{tabular}{|ccccc|}
Sample & $\log_{10}(Z_{0T}\ [Z_{\rm Fe,\odot}])$ & $\beta_T$ & $\sigma\log_{10}Z_{\rm Fe}|T$ & $r_s$ \\
\hline
AGN full & $-0.178\pm0.011$ & $-0.10\pm0.04$ & $0.16\pm0.01$ & $-0.21\ (2\times10^{-3})$ \\
AGN NCC & $-0.190\pm0.013$ & $-0.14\pm0.04$ & $0.16\pm0.01$ & $-0.27\ (4\times10^{-4})$ \\
AGN CC & $-0.157\pm0.023$& $0.00\pm0.07$ & $0.15\pm0.02$ & $-0.08\ (6\times10^{-1})$ \\
CSF ($T<1.7$ keV) & $0.035\pm0.039$& $0.33\pm0.15$ & $0.19\pm0.01$ & $0.21\ (5\times10^{-3})$ \\
CSF ($T\ge1.7$ keV) & $0.082\pm0.032$& $-0.06\pm0.09$ & $0.14\pm0.01$ & $-0.04\ (8\times10^{-1})$ \\
CHEERS data & $-0.127\pm0.019$ & $0.00\pm0.06$ & $0.11\pm0.01$ & $+0.03\ (8\times10^{-1})$
 \end{tabular}

 \end{center}
 \end{table*}
%

The AGN and CHEERS samples show almost no correlation between the iron abundance and the temperature in the cluster core. The Spearman's correlation coefficients obtained from simulations reveal either no correlation ($|r_s| <0.1$ with 60-80 per cent probability of null hypothesis) or a very low-level of correlation ($|r_s| \sim 0.2-0.3$) with null-hypothesis probability of order of $10^{-3}$-$10^{-4}$ . The CHEERS data shows a flat distribution of iron abundance across the temperature range with the slope $\beta_T$ consistent with zero within the uncertainty. Also, in this case the two quantities are uncorrelated ($r_s\sim 0$) at higher significance (80 per cent consistent with no correlation).
The relation for our AGN data also presents a very shallow slope
$\beta_{T}\sim-0.1$, which implies that, on average, a cluster that is
10 times hotter than a group has only about $20\%$ lower
metallicity. This indicates that the ICM iron abundance is
statistically constant from groups to clusters in the entire sample. On the other hand, the CSF run exhibits a flat distribution of iron abundance only in high-temperature systems ($T>1.7$ keV), while in the low-temperature regime the CSF $Z_{\rm Fe}$ falls as temperature decreases with the slope $\beta_T\sim0.3$.
  \begin{figure}
\centering
{\includegraphics[width=0.45\textwidth]{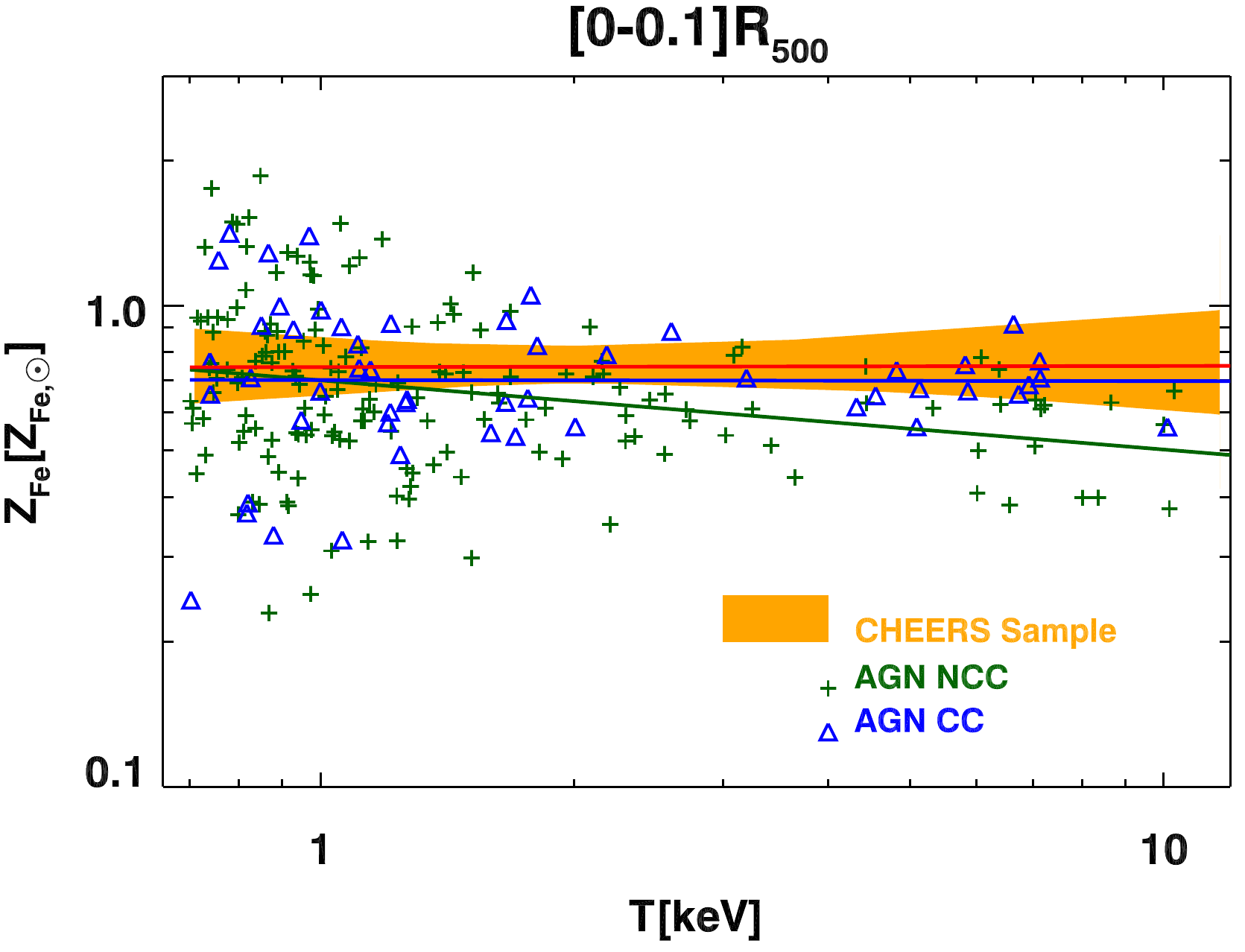}}
\caption{$Z_{\rm Fe}-T$ relation for simulated CC (blue triangles) and NCC (green crosses) subsamples obtained from the AGN simulation shown along with their best-fit relations (solid lines). The solid red line and the orange-shaded area are the best-fit relation and the $68.3\%$ confidence region, respectively, for the CHEERS sample as shown in Fig.~\ref{fig2}.}
   \label{fig3}
\end{figure}

Compared to the CHEERS sample, the iron abundance of the AGN simulated objects with $T\geq 2$ keV 
is slightly lower than the observed value at a given temperature. At
the pivot point of temperature $1.7$ keV, the observed $Z_{\rm Fe}$ is
higher than the simulated one by $13$ per cent, yet the two values are
consistent within the intrinsic scatters. While, the CSF $Z_{\rm Fe}$ is about $70\%$ higher than the CHEERS value at the pivot temperature. The offset in
normalisation between simulations and observations can be ascribed to several factors including the sample selection as discussed in the following section and the
details of the model for chemical enrichment used in the simulations,
e.g., the stellar initial mass function and the sets of yields (see discussion in Section 5.1). Therefore, the slight offset between simulated and observed data should be not considered in itself a reason of serious concern. It is, however, relevant to remark that the offset between the AGN and the CSF samples reflects an intrinsic difference in the metal production predicted by these two models since they are both obtained with the same description of stellar and chemical evolution. %
\vspace{0.5cm}

We note that the simulated data exhibit slightly higher scatter (e.g., in AGN simulation, $\sigma\log_{10}Z_{\rm Fe}|T=0.16$) than the observed CHEERS data ($\sigma\log_{10}Z_{\rm Fe}|T=0.11$). It is evident from Fig.~\ref{fig2} that the larger scatter in simulated data is mostly due to the statistics of  low-temperature groups ($T<1.7$ keV). It is important to notice that in this low-temperature range, there is about $70$ per cent of the AGN objects, while the number of low-temperature objects in the CHEERS sample is $\sim42$ per cent. Observationally, it is more challenging to both detect and observe groups rather than clusters, particularly due to the depth required to obtain a good measurement of temperature and metal abundances. The observational X-ray selected sample thus tends to favor the inclusion of massive cool core clusters that are easier to detect and observe, whereas simulations have a larger fraction of groups. This is mostly due to the steepness of the mass function that enhances the statistics of smaller objects in quite large volume-limited zoom-in regions\footnote{We note that the simulated sample is not strictly volume-limited as the objects are obtained from the high-resolution zoomed-in regions. However, as each region's radius extends the size of at least 5 times $R_{vir}$ of the central object, this volume is large enough to contain several other groups.}. These low-mass simulated systems have small radius (the mean $R_{500}\sim450$ kpc) so that the level of metallicity in the central region ($<0.1R_{500}$) is very sensitive to AGN feedback, the efficiency of radiative cooling, and the dynamical state of the system. To conclude, the larger scatter found in simulations compared to observations is mostly due to the CHEERS small statistics of low-temperature systems ($T<1.7$ keV). Further contributions may come from the observational difficulties of detecting groups with a low-metallicity core since they are likely less peaked in the X-rays.


\subsubsection{Cool-core and non-cool-core $Z_{\rm Fe}-T$ relations}

Cool-core clusters are shown to host a larger amount of metals in the core region compared to non cool-core clusters (e.g, \citealt{DeGrandi.etal.2004}). Therefore we might expect a difference in the $Z_{\rm Fe}-T$ relations derived from each separate population that needs to be evaluated to estimate any potential bias in observed samples. This is particularly relevant for the CHEERS sample as its member clusters are selected by RGS (the Reflection Grating Spectrometer onboard {\it XMM-Newton}) which is selectively sensitive to centrally-peaked clusters (\citealt{de_plaa.etal.2017}) and peakiness can be considered as a ``proxy'' for the cool-coreness (e.g., \citealt{mantz.etal.2017}). To address this issue, we divide the simulated AGN dataset into cool-core and non cool-core subsamples, and study the iron abundance-temperature relation for each of them separately. The same task cannot be done for the CSF dataset since in our CSF simulation, the diversity of CC and NCC systems is not present due to the lack of an efficient form of feedback, e.g., feedback from AGN (see \citealt{rasia.etal.2015} and \citealt{biffi.etal.2017} for detailed discussions). For the selection of simulated CC systems, we compute the pseudo-entropy defined as: 
\begin{equation}
\sigma=\frac{(T_{\rm IN}/T_{\rm OUT})}{(EM_{\rm IN}/EM_{\rm OUT})^{1/3}}, \label{eq5}
\end{equation}
where $T$ is the spectroscopic-like temperature and $EM$ is the emission measure, computed in the IN ($r<0.05\times R_{180}$) and OUT ($0.05\times R_{180}<r<0.2\times R_{180}$) regions. Following \cite{rasia.etal.2015} we define the CC systems as those that have a pseudo-entropy value lower than $0.55$, whereas systems with $\sigma > 0.55$ are classified as NCC. This definition of CC clusters was originally introduced to compare the statistics of CCs in the simulated sample of \cite{rasia.etal.2015} with the observational ratio found by \cite{rossetti.etal.2011}. We keep the same criterion here because it appropriately mimics that of the CHEERS sample for the following two reasons:
\begin{itemize}
    \item The pseudo-entropy depends on both the emission-measure ratio and the temperature ratio. However, the former is the dominant quantity as CCs have a steeply declining emission profiles with radius. This propriety is common with the CHEERS clusters that, indeed, have been selected for having a clear X-ray peak in their core. 
    \item Our AGN simulations exhibit a strong correlation between the pseudo-entropy and the central entropy, namely all simulated CCs have low pseudo-entropy (by definition) and at the same time they all have a low value of the central entropy. Similarly, also the CHEERS objects are characterised by a low central entropy (below $30\ \rm{keV}\rm{cm}^2$ accordingly to table A.2 in \citealt{pinto.etal.2015}).   
\end{itemize}
For these reasons, a cool-core object is similarly identified in our simulations and in the observational sample that we compare with. Therefore, we do not investigate other criteria used in literature (\citealt{cavagnolo.etal.2009, hudson.etal.2010,leccardi.etal.2010, McDonald.etal.2013,pascut.ponman.2015,barnes.etal.2018}).

We show in Fig.~\ref{fig3} the $Z_{\rm Fe}-T$ relation for both the CC and NCC subsamples obtained from the AGN simulation along with the CHEERS data. In general the CC systems have slightly higher central iron abundance, e.g., $\sim8\%$ at 1.7 keV, than the NCC ones, which is consistent with recent numerical and observational studies showing that the ICM iron profile in CC clusters is steeper and peaked in the central regions (\citealt{leccardi.etal.2010,rasia.etal.2015, ettori.etal.2015, biffi.etal.2017}). Whereas the difference is more prominent for clusters with $T>2$ keV, for the low-temperature systems, the separation between the metallicity level in CCs and NCCs is not well-established and the scatter is large. As a consequence, the slope of the $Z_{\rm Fe}-T$ of CCs is flatter than that of NCCs. Hence, for the study of $Z_{\rm Fe}-T$ relation, the effect of solely including CC systems mainly affects the slope of the high-temperature end, whereas the effect is not
very significant in the low-T regime.
  \begin{figure*}
\centering
\begin{center}
{\includegraphics[width=\textwidth]{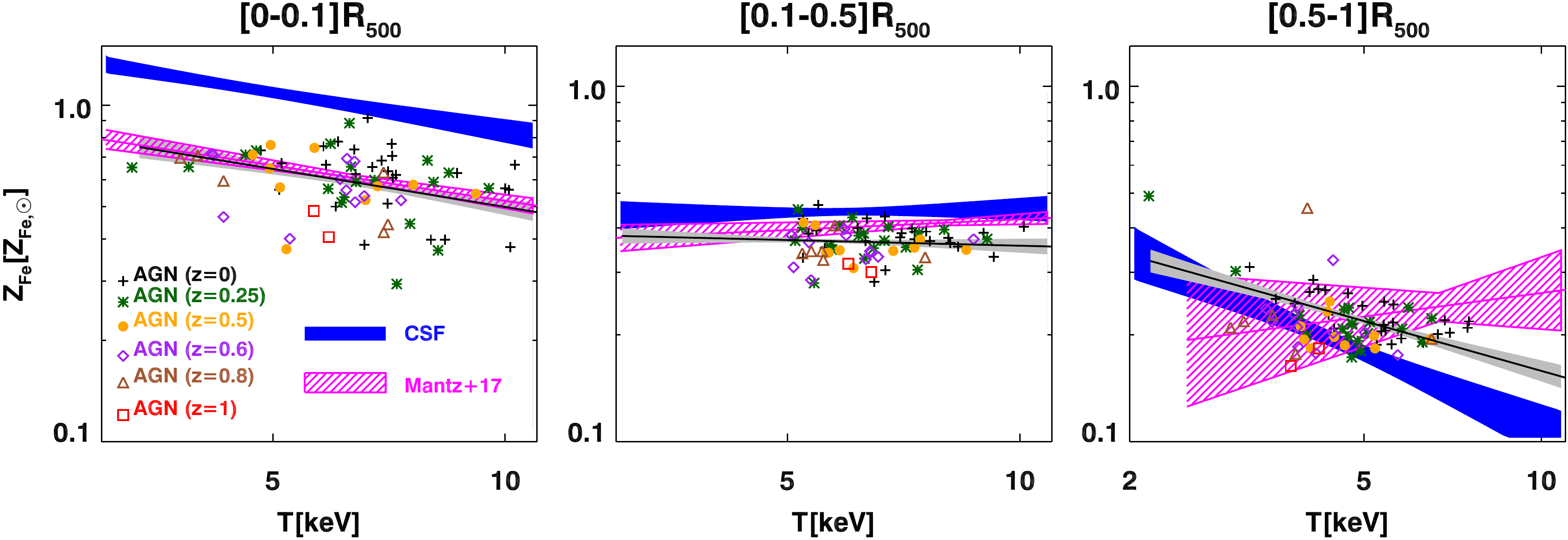}}
\end{center}
\caption{Comparison between the simulated $Z_{\rm Fe}-T$ relations and the observational results from \citealt{mantz.etal.2017} within different radial ranges for massive systems ($T_{[0.1-0.5]R_{500}}>5$ keV). The AGN best-fit relation is represented with a black solid line and the grey-shaded area specifies the $68.3\%$ confidence region, while the blue-shaded area represents the corresponding confidence region for the CSF simulation. The observational constraint region, which is confined between the two magenta curves, is derived based on the best-fit values of normalisation, slope, and their associated $1 \sigma$ uncertainties (see Table~\ref{tb2}). The simulated and observed best-fit relations are evaluated at the pivot redshift and temperature as reported in Table~\ref{tb2}.}
 \label{fig4}
\end{figure*}

The CC subsample is in better agreement with the CHEERS dataset than the NCC subsample, not only in terms of the slope, $\beta_T\sim0$, but also in terms of the normalisation. The simulated and observed normalisations are consistent at $1\sigma$ and the relative offset is less than 0.05 $Z_{\rm Fe,\odot}$. The cores of both simulated CC and observed clusters present a constant iron abundance equal to $Z_{\rm Fe}\simeq0.75 Z_{\rm Fe,\odot}$ over the considered range of temperature  with a dispersion of $\sim40$ per cent and $\sim30$ per cent, respectively. However, we remind that the absolute value of the metallicity-temperature normalisation depends on metal enrichment model and its assumptions (discussed in more detail in Section 5.1), therefore the matching (or mismatching) value of the normalisations for simulations and observations should be not over-interpreted. Instead, it is worth stressing how the simulated CC subsample is in better agreement with the CHEERS objects.

\begin{figure*}
\begin{center}
{\includegraphics[width=\textwidth]{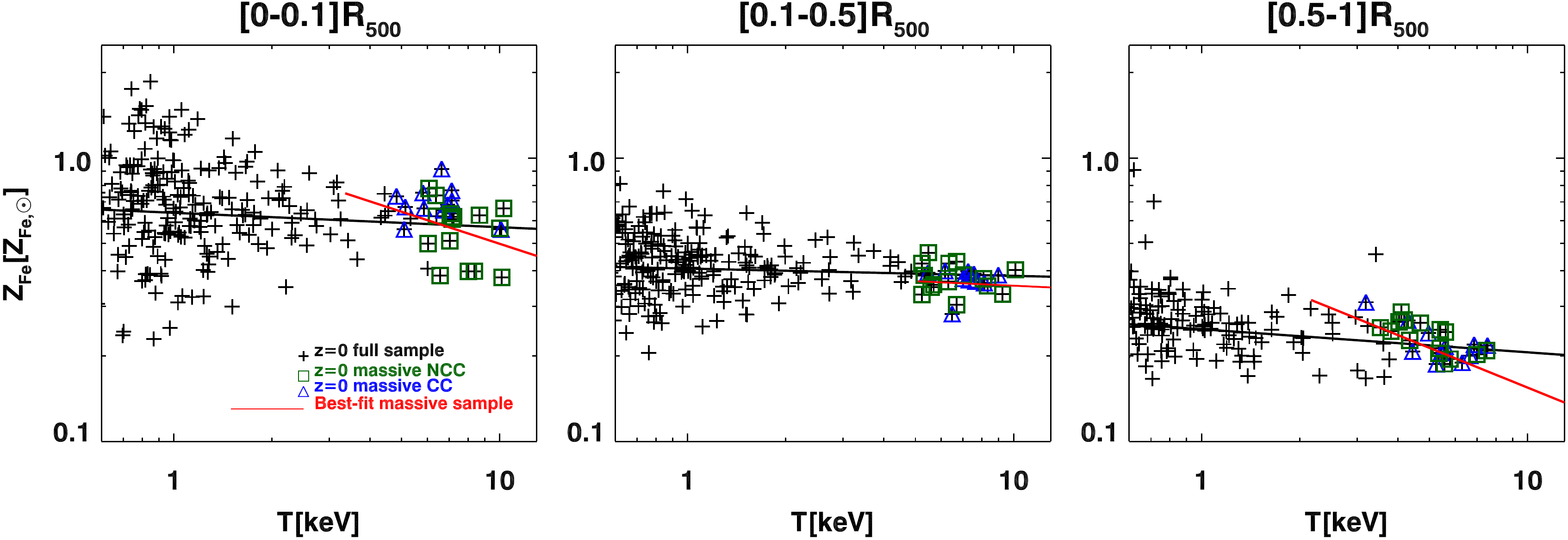}}
\end{center}
\caption{Comparison between the AGN $Z_{\rm Fe}-T$ simulated relations at $z=0$ derived from the entire sample and the massive subsample. The green squares (NCC) and blue triangles (CC) specify massive clusters that are selected according to their temperature $T_{[0.1-0.5]R_{500}} >5$ keV. The solid lines are best-fitting relations corresponding to the whole sample at $z=0$ (black) and to the selected massive sample from $z=0$ to $z=1$ (red).}
 \label{fig5}
\end{figure*}
\subsection{$Z_{\rm Fe}-T$ evolution in massive clusters}

In this subsection we compare the AGN and CSF simulated $Z_{\rm Fe}-T$ relations with the results obtained by \cite{mantz.etal.2017} for massive clusters.
To be consistent with them, we select simulated clusters with $T_{[0.1-0.5]R_{500}}>5$ keV, where $T_{[0.1-0.5]R_{500}}$ is the temperature
measured within the region $[0.1-0.5]R_{500}$. We consider 6 different snapshots corresponding to redshifts:
$z=0, 0.25, 0.5, 0.6, 0.8$, and $1$, where we, respectively, identify $26(30), 19(23), 10(15), 11(15), 6(8),$ and $2(3)$ objects in the AGN (CSF) simulation.
 For this comparison, we use the projected emission-weighted iron abundance and the spectroscopic-like temperature computed within three
 radial ranges: $[0-0.1]R_{500}$, $[0.1-0.5]R_{500}$, and $[0.5-1]R_{500}$.
 Following the approach by Mantz et al., we quantify the correlation between $Z_{\rm Fe}$ and $T$ in addition to study the evolution of the relation by simultaneously fitting all simulated data with the formula: 
\begin{equation}
Z_{\rm Fe}=Z_{0T}\times\bigg(\frac{1+z}{1+z_{\rm{piv}}}\bigg)^{\gamma_z}\times\bigg(\frac{T}{T_{\rm{piv}}}\bigg)^{\beta_T},\label{eq7}
\end{equation}
where the free parameters are: the normalisation $Z_{0T}$, the slope $\beta_T$, and the redshift evolution of the iron abundance, $\gamma_{z}$. We fix the pivot values of redshift and temperature, $z_{\rm{piv}}$ and $T_{\rm{piv}}$, to the same values used in the observational analysis reported in Table~\ref{tb2}, where we also list the best-fitting parameters of the $Z_{\rm Fe}-T$ relation from our analysis as well as from the work of \cite{mantz.etal.2017} (see their Table~1).
\begin{table*}
 \caption{\label{tb2}
  Best-fitting parameters of the relation in Eq. (\ref{eq7}) for the AGN simulation shown along with the results from \citealt{mantz.etal.2017}. The Spearman's rank correlation coefficients ($r_s$) for the AGN simulation and the null hypothesis probability (in parentheses) are also reported (these quantities are not available, N/A, for the observational data).}
 \begin{center}
 \begin{tabular}{c|ccccc|cccc}
 \hline
$Aperture$ & $z_{\rm{piv}}$ & $T_{\rm{piv}}[\rm{keV}]$ & $\log_{10}(Z_{0T}\ [Z_{\rm Fe,\odot}])$ & $\beta_{\rm{T}}$ & $\gamma_{\rm{z}}$ &$\sigma_{\log_{10}Z_{\rm Fe}|T}$ & $r_s$ \\
 \hline
 {\bf AGN} \\
 $[0.0-0.1]R_{500}:$ & $0.23$ & $6.4$ &$-0.230\pm0.011$ & $-0.38\pm0.11$ & $-0.32\pm0.13$ & $0.09\pm0.01$ & $-0.40\ (4\times10^{-4})$ \\
 $[0.1-0.5]R_{500}:$ & $0.19$ & $8.0$ & $-0.440\pm0.009$ & $-0.05\pm0.08$ & $-0.18\pm0.06$ & $0.045\pm0.004$ & $+0.01\ (9\times10^{-1})$\\
 $[0.5-1]R_{500}:$ & $0.17$ & $6.7$ & $-0.720\pm0.016$ & $-0.47\pm0.09$ & $-0.34\pm0.10$ & $0.07\pm0.01$& $-0.38\ (9\times10^{-4})$\\
 {\bf CSF} \\
 $[0.0-0.1]R_{500}:$ & $0.23$ & $6.4$ &$-0.003\pm0.021$ & $-0.38\pm0.10$ & $-0.46\pm0.17$ & $0.16\pm0.01$ & $-0.30\ (4\times10^{-3})$ \\
 $[0.1-0.5]R_{500}:$ & $0.19$ & $8.0$ & $-0.350\pm0.014$ & $0.03\pm0.13$ & $-0.37\pm0.12$ & $0.10\pm0.01$ & $+0.17\ (1\times10^{-1})$\\
 $[0.5-1]R_{500}:$ & $0.17$ & $6.7$ & $-0.843\pm0.041$ & $-0.72\pm0.20$ & $-0.93\pm0.24$ & $0.19\pm0.01$& $-0.04\ (7\times10^{-1})$\\
 {\bf \cite{mantz.etal.2017}} \\
  $[0.0-0.1]R_{500}:$ & $0.23$ & $6.4$ &$-0.217\pm0.009$ & $-0.35\pm0.06$ & $-0.14\pm0.17$ & $0.08\pm0.01$ & N/A \\
 $[0.1-0.5]R_{500}:$ & $0.19$ & $8.0$ & $-0.384\pm0.007$ & $0.10\pm0.07$ & $-0.71\pm0.15$ & $0.04\pm0.01$ & N/A\\
 $[0.5-1]R_{500}:$ & $0.17$ & $6.7$ & $-0.622\pm0.040$ & $0.22\pm0.34$ & $-0.30\pm0.91$ & $0.00^{+0.07}_{-0.00}$ & N/A\\
  \end{tabular}
 \end{center}
 \end{table*}
As shown in Fig.~\ref{fig4}, the AGN simulated radial trend of the averaged iron abundance is consistent with the data showing a decrease of similar amplitude from the cluster core to the outskirts. On average, both simulated and observed iron abundance are higher in the innermost region, where $Z_{\rm Fe}\sim0.6 Z_{\rm Fe,\odot}$, and they gradually decrease to the level of $0.2 Z_{\rm Fe,\odot}$ in the most external radial range. This result is in line with previous works on ICM metallicity profiles from both simulations and observations (e.g., \citealt{werner.etal.2013, rasia.etal.2015, urban.etal.2017,biffi.etal.2017, biffi.etal.2018}). On the other hand, the CSF simulated clusters are largely inconsistent with the data: the iron abundance is about $70$ per cent higher than the observed value in the central region ($r<0.1R_{500}$), but rapidly decreasing with temperature in the most external regions ($r>0.5R_{500}$).


The AGN simulation also agrees well with the analysis by Mantz et al. on the $Z_{\rm Fe}-T$ intrinsic scatter. Among the three considered radial ranges, both simulations and observations present the largest scatter in the core (see Table~\ref{tb2}), that is strongly affected by several astrophysical processes such as feedback from the AGN, or intense stellar activity. The CSF clusters exhibit higher scatter among the three considered ranges in comparison to the AGN and Mantz et al. results.

In terms of the slopes, we notice a good agreement between both simulations and observations in the two regions within half of $R_{500}$, while the best-fitting lines seem to have opposite trends in the cluster outskirts ($\beta_{\rm AGN}=-0.47$, $\beta_{\rm CSF}=-0.72$, and $\beta_{\rm obs}=0.22$). However, the observational constraints are weak and considering the 1$\sigma$ error associated with $\beta_{\rm obs}$, equal to 0.34, the AGN and observational slopes are consistent within $2\sigma$. In addition, we notice that most of the AGN simulated points are well within the shaded area which shows the observational 1$\sigma$ dispersion around the best-fitting line.

It is interesting to note that the pronounced steepness of the AGN data is caused by the specific selection of the hottest clusters as  illustrated in Fig.~\ref{fig5}, where the best-fitting relation of the most massive sample (red line) is compared with the overall trend (black line). In particular, the steep relation that characterises the core of the massive simulated clusters is biased by the
segregation between the CC and NCC systems: the CCs have both higher metallicity and lower temperature compared to the NCCs. This separation disappears when the groups are added to the sample (see also Fig.~\ref{fig3}). According to \cite{barnes.etal.2018}, in their simulations this result might depend on the operational definition of cool-core systems. Addressing this issue with our simulated clusters is nevertheless beyond the scope of this work.

Finally, the evolution of the simulated $Z_{\rm Fe}-T$ relations is positive in all radial ranges: at fixed temperature, the iron abundance increases with time (see $\gamma_z$ values in Table~\ref{tb2}). The amplitude of this variation in the AGN simulation is however limited, as the iron abundance on average grows by less than 30\% from $z=1$ to $z=0$. This result is consistent with the results from Mantz et al. (except in the intermediate range) and from other observational analyses (\citealt{ettori.etal.2015,McDonald.etal.2016}). Differently, the CSF iron abundance in general evolves more rapidly in comparison to the observational data, except in the intermediate range $[0.1-0.5]R_{500}$, due to highly efficient star formation in the CSF simulation.
    

As for the radial trend of the evolution, the simulated data show that the normalisation of the $Z_{\rm Fe}-T$ relation evolves almost equally in the three radial ranges, except for the enhancement of the CSF $Z_{\rm Fe}$ in the range $[0.5-1]R_{500}$, while the observed data by \cite{mantz.etal.2017} exhibit a stronger evolution of the iron abundance, with $\gamma_{z}=-0.71\pm0.15$, in the intermediate radial range. The authors suggest that the late-time increase of iron abundance of the gas in the intermediate radial range could be due to the mixing with enriched gas from the cluster centres caused by mergers or AGN outflows. Observational results on the spatial pattern of metallicity evolution is, however, still matter of debate. At variance with the conclusions by \cite{mantz.etal.2017}, \cite{ettori.etal.2015} show that the ICM metallicity slightly evolves in the central region of the CC clusters only and, even in these objects, it remains constant at larger radii. On the other hand, \cite{McDonald.etal.2016} point out that the ICM metallicity is consistent with no evolution outside of the core.

\begin{figure*}
\begin{center}
{\includegraphics[width=\textwidth]{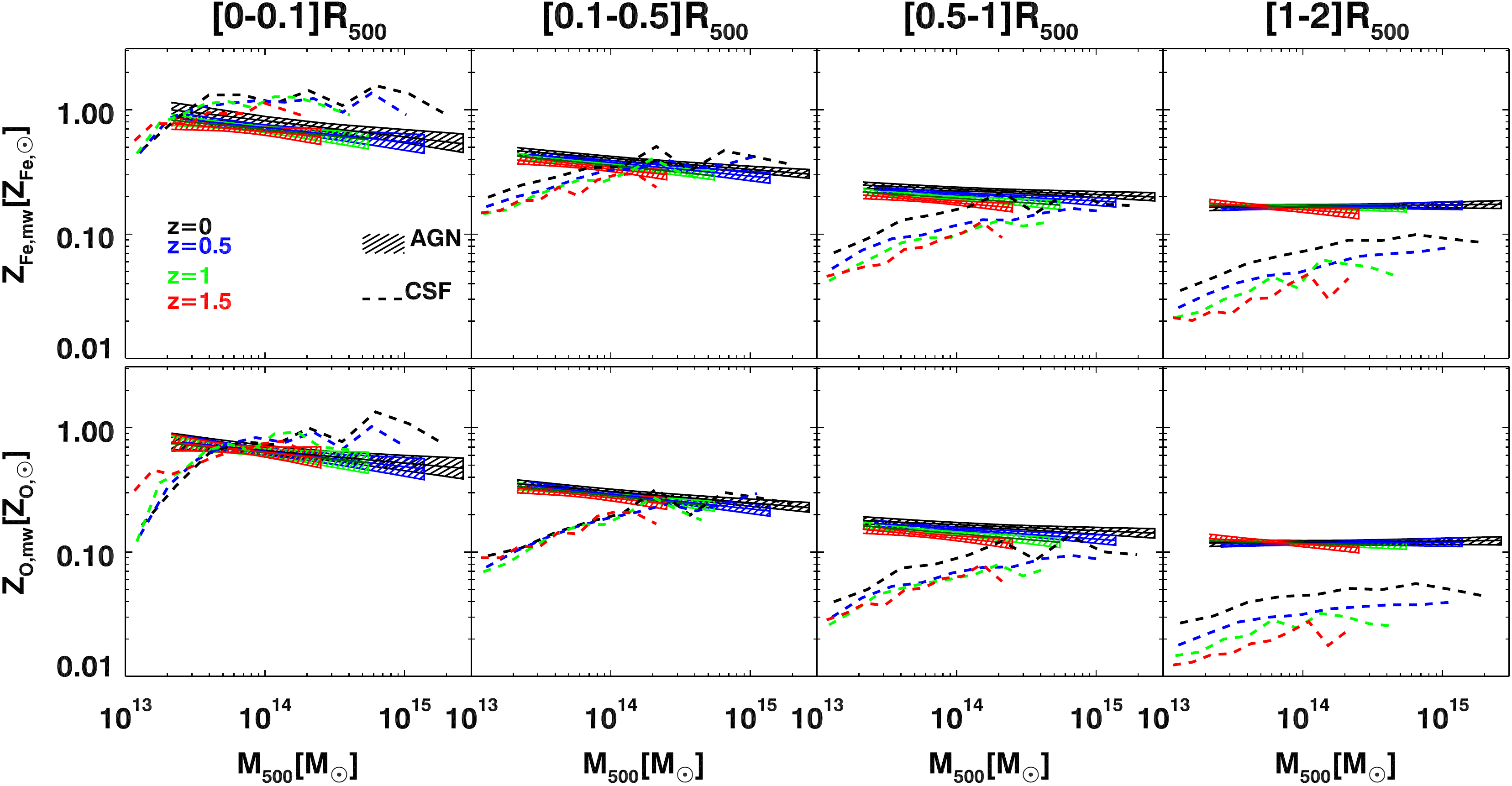}}
\end{center}
\caption{The relation between cluster mass and mass-weighted iron abundance (upper) and oxygen abundance (lower), at different ranges of radius and redshift are shown for AGN and CSF simulations. Solid lines are the AGN best-fit relations as described by Eq.~(\ref{eq2}) with the parameters reported in Table~\ref{tb3}, while the shaded areas specify the $68.3\%$ confidence regions. The dashed lines represents median relations obtained from the CSF run.}
 \label{fig6}
\end{figure*}
\begin{figure*}
\begin{center}
{\includegraphics[width=\textwidth]{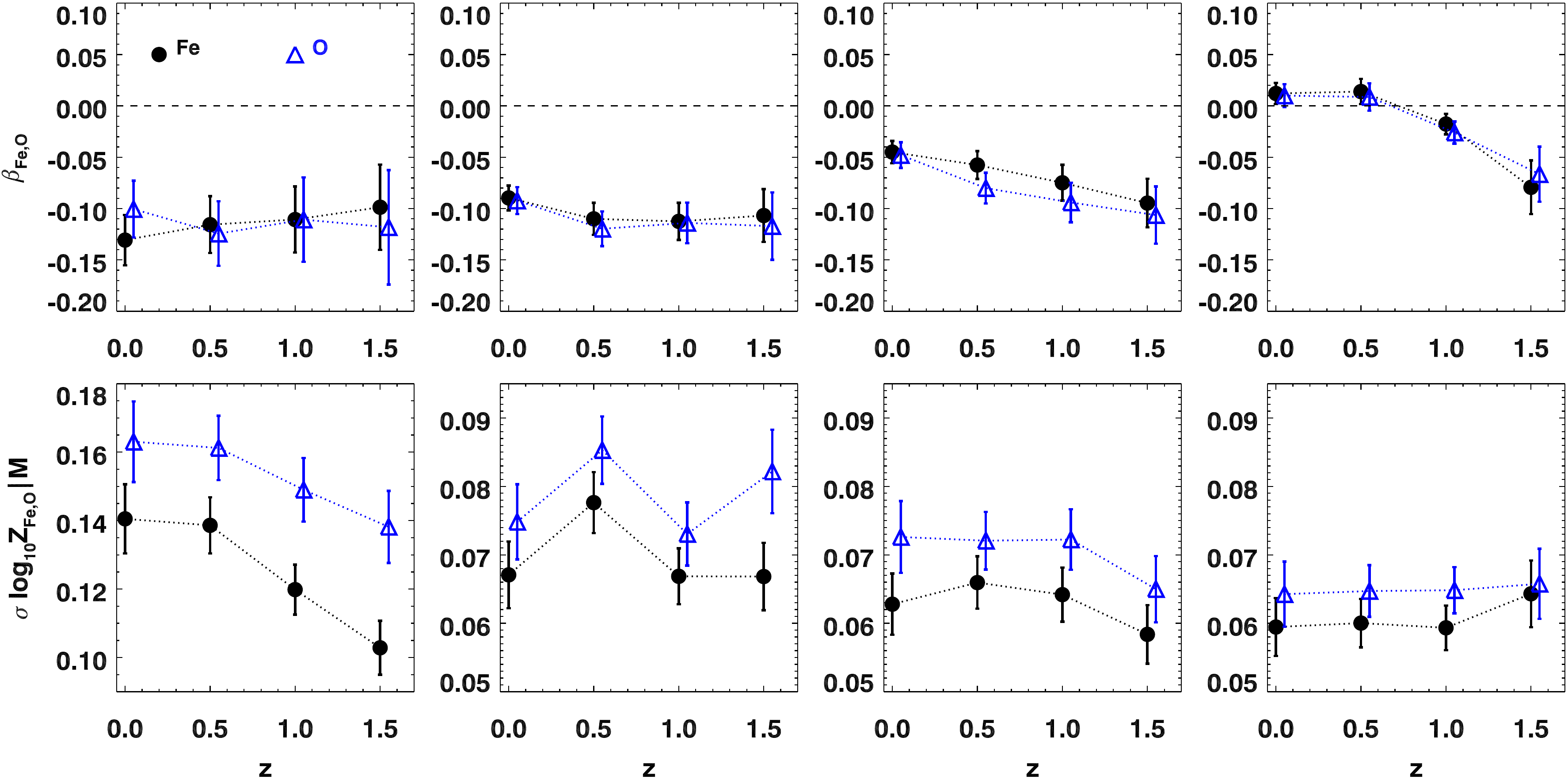}}
\end{center}
\caption{The slope (upper) and scatter (lower) of the $Z_{\rm Fe}-M_{500}$ and $Z_{\rm O}-M_{500}$ relations derived from the AGN simulations as a function of redshift for the four considered radial ranges, from left to right, $[0-0.1]R_{500}$, $[0.1-0.5]R_{500}$, $[0.5-1]R_{500}$, and $[1-2]R_{500}$, respectively. The error bars quote $1\sigma$ uncertainty of the best-fit slopes and the best-fit intrinsic scatters of the relations as described by Eq.~(\ref{eq2}).}
 \label{fig7}
\end{figure*}
\section{Mass-Metallicity Relation and Evolution}
\label{Mass_Metallicity Relation and Evolution}
  \begin{table*}
 \caption{\label{tb3}
  Best-fitting parameters of Eq.~(\ref{eq2}) for the $Z_{\rm Fe}-M_{500}$ and $Z_{\rm O}-M_{500}$ relations in the AGN simulation, shown along with Spearman's correlation coefficients and the null hypothesis probability (in the parentheses).}
 \begin{center}
 \resizebox{\textwidth}{!}{
 \begin{tabular}{|c|c|ccc|cccc}
 \hline
 AGN & & Fe & & & & O  & \\
 \hline
$z$ & $\log_{10}(Z_{\rm 0,Fe}\ [Z_\odot])$ & $\beta_{\rm Fe}$ & $\sigma_{\log_{10}Z_{{\rm Fe}}|M}$ & $r_{s,{\rm Fe}}$ & $\log_{10}(Z_{\rm 0,O}\ [Z_\odot])$ & $\beta_{\rm O}$ & $\sigma_{\log_{10}Z_{\rm O}|M}$ & $r_{s,{\rm O}}$\\
 \hline
 $[0.0-0.1]R_{500}:$ \\
0.0 &  $-0.088\pm0.015$  & $-0.131\pm0.025$  &$0.141\pm0.010$&$-0.49\ (1\times10^{-7})$ & $-0.183\pm0.018$  & $-0.101\pm0.028$  &$0.163\pm0.012$ &$-0.39\ (3\times10^{-5})$ \\
0.5 &  $-0.143\pm0.011$  & $-0.116\pm0.028$  &$0.139\pm0.008$& $-0.30\ (2\times10^{-4})$ &$-0.189\pm0.013$  & $-0.124\pm0.031$  &$0.161\pm0.009$ & $-0.30\ (2\times10^{-4})$\\
1.0 &  $-0.158\pm0.011$  & $-0.111\pm0.032$  &$0.120\pm0.007$& $-0.29\ (6\times10^{-4})$ & $-0.200\pm0.014$  & $-0.111\pm0.041$  &$0.149\pm0.009$& $-0.19\ (3\times10^{-2})$\\
1.5 &  $-0.175\pm0.014$  & $-0.099\pm0.042$  &$0.103\pm0.008$& $-0.28\ (5\times10^{-3})$ & $-0.196\pm0.018$  & $-0.118\pm0.056$  &$0.138\pm0.011$ & $-0.24\ (2\times10^{-2})$\\
 \hline
  $[0.1-0.5]R_{500}:$ \\
0.0 &  $-0.390\pm0.007$  & $-0.090\pm0.012$  &$0.067\pm0.005$&$-0.60\ (2\times10^{-11})$  & $-0.512\pm0.008$  & $-0.092\pm0.013$  &$0.075\pm0.005$ & $-0.56\ (7\times10^{-10})$\\
0.5 &  $-0.424\pm0.006$  & $-0.110\pm0.016$  &$0.078\pm0.004$& $-0.50\ (1\times10^{-11})$ & $-0.533\pm0.007$  & $-0.120\pm0.017$  &$0.085\pm0.005$ &$-0.51\ (1\times10^{-11})$\\
1.0 &  $-0.447\pm0.006$  & $-0.112\pm0.018$  &$0.067\pm0.004$& $-0.42\ (3\times10^{-7})$ & $-0.547\pm0.007$  & $-0.114\pm0.020$  &$0.073\pm0.005$ & $-0.41\ (8\times10^{-7})$\\
1.5 &  $-0.478\pm0.009$  & $-0.107\pm0.026$  &$0.067\pm0.005$& $-0.37\ (2\times10^{-4})$ & $-0.564\pm0.011$  & $-0.117\pm0.033$  &$0.082\pm0.006$ & $-0.31\ (2\times10^{-3})$\\
  \hline
    $[0.5-1.0]R_{500}:$ \\
0.0 &  $-0.637\pm0.007$  & $-0.045\pm0.011$  &$0.063\pm0.004$& $-0.37\ (1\times10^{-4})$ & $-0.780\pm0.008$  & $-0.048\pm0.012$  &$0.073\pm0.005$ & $-0.37\ (1\times10^{-4})$\\
0.5 &  $-0.676\pm0.005$  & $-0.058\pm0.013$  &$0.066\pm0.004$& $-0.35\ (6\times10^{-6})$ & $-0.816\pm0.006$  & $-0.080\pm0.015$  &$0.072\pm0.004$ &$-0.40\ (1\times10^{-7})$\\
1.0 &  $-0.711\pm0.006$  & $-0.075\pm0.017$  &$0.064\pm0.004$& $-0.34\ (4\times10^{-5})$ & $-0.852\pm0.007$  & $-0.094\pm0.019$  &$0.072\pm0.004$ &$-0.40\ (1\times10^{-6})$\\
1.5 &  $-0.749\pm0.008$  & $-0.095\pm0.024$  &$0.058\pm0.004$& $-0.37\ (2\times10^{-4})$ & $-0.889\pm0.009$  & $-0.106\pm0.028$  &$0.065\pm0.005$ &$-0.39\ (6\times10^{-5})$\\
\hline
$[1.0-2.0]R_{500}:$ \\
0.0 &  $-0.778\pm0.006$  & $ 0.012\pm0.011$  &$0.059\pm0.004$& $+0.09\ (4\times10^{-1})$ & $-0.921\pm0.007$  & $ 0.008\pm0.011$  &$0.065\pm0.005$ & $+0.07\ (5\times10^{-1})$\\
0.5 &  $-0.782\pm0.005$  & $ 0.012\pm0.012$  &$0.060\pm0.003$& $+0.05\ (5\times10^{-1}) $& $-0.930\pm0.005$  & $ 0.011\pm0.013$  &$0.065\pm0.004$ & $+0.04\ (7\times10^{-1})$\\
1.0 &  $-0.781\pm0.006$  & $-0.052\pm0.018$  &$0.066\pm0.004$& $-0.20\ (2\times10^{-2})$ & $-0.928\pm0.007$  & $-0.040\pm0.021$  &$0.074\pm0.004$ & $-0.14\ (1\times10^{-1})$\\
1.5 &  $-0.805\pm0.009$  & $-0.085\pm0.025$  &$0.065\pm0.005$& $-0.27\ (7\times10^{-3})$ & $-0.937\pm0.009$  & $-0.067\pm0.027$  &$0.066\pm0.005$ & $-0.21\ (3\times10^{-2})$\\
  \end{tabular}}
 \end{center}
 \end{table*}

 After verifying that our numerical model generally reproduces observational findings, we provide here a detailed prediction on the scale invariance of the metallicity distribution and its evolution. In observational data the ICM temperature is used as mass-proxy, however, since both temperature and metallicity are derived from the same spectra there is a certain degeneracy between the two quantities. Furthermore, the  measurements of the temperature in the central regions can be biased low because of multi-temperature gas. In the previous section we mimic the spectral measurements of metallicity and temperature by using projected quantities and we showed agreement between simulated and observed data. In this section, we take advantage of the precise knowledge of the mass from simulations and we study how the distribution and evolution of metallicity depends on the total mass ($M_{500}$) and how this trend is influenced by AGN feedback. At first, we will analyse the mass dependence of both iron and oxygen in AGN and CSF simulations, while in the second part we will focus exclusively on the iron because the two metal elements present very similar behaviours. All relations are extended to poor groups, $M_{500} > 1\times 10^{13} M_{\odot}$, satisfying the condition reported at the end of Section~2.1.

\subsection{The $Z_{\rm Fe}-M_{500}$ and $Z_{\rm O}-M_{500}$ relations}
 The relationships between $Z_{\rm Fe}$, $Z_{\rm O}$ and $M_{500}$ are computed within the same apertures as before with the addition of one more external region: $[1-2]R_{500}$. The relations are evaluated separately at four fixed redshifts: $z=0,\ 0.5,\ 1,$ and $1.5$. At a second stage, we quantify the evolution of the relation combining the samples. 
\begin{figure*}
\begin{center}
{\includegraphics[width=\textwidth]{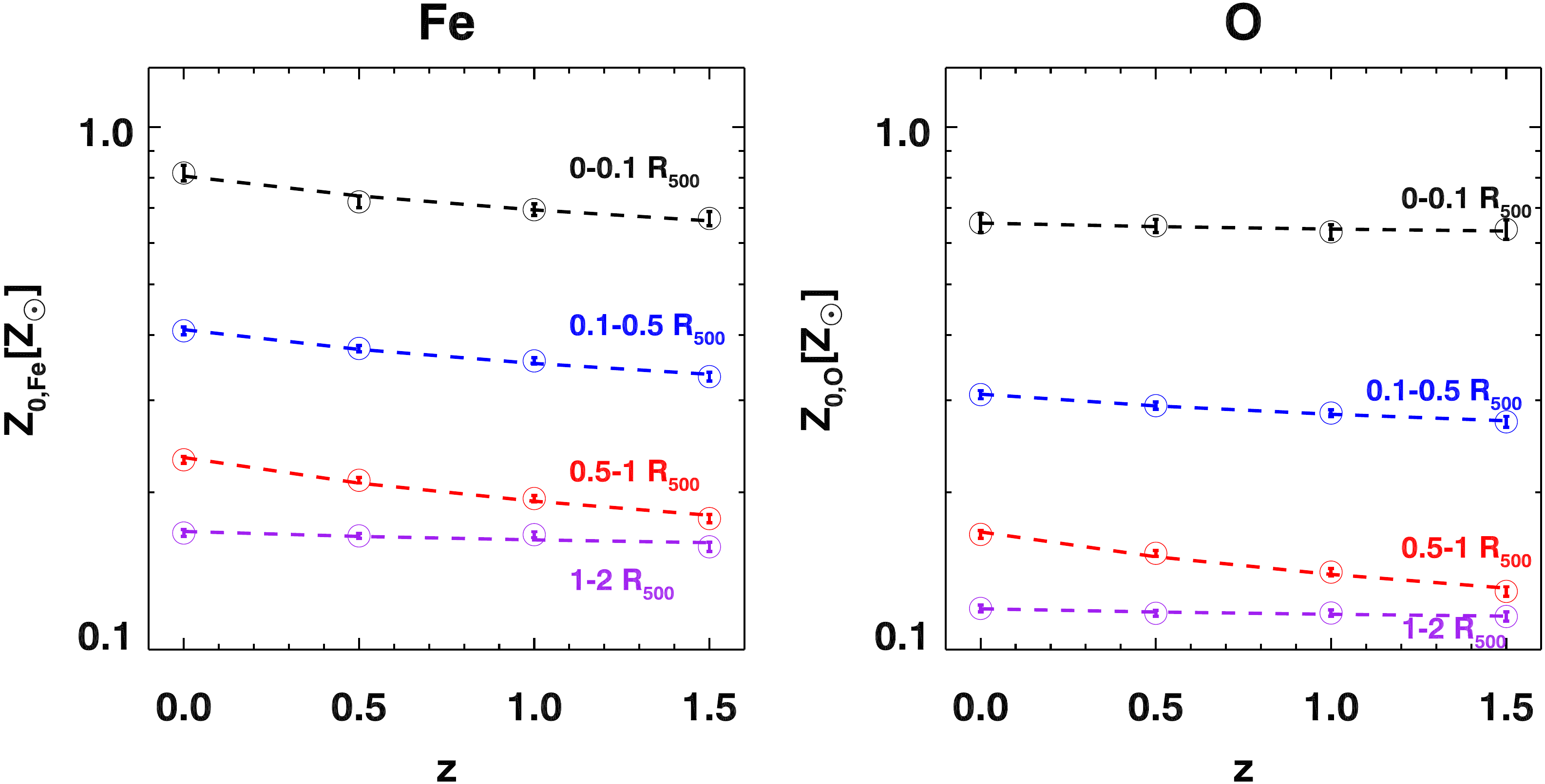}}
\end{center}
\caption{Evolution of the normalisation of the AGN simulated $Z_{\rm Fe}-M_{500}$ (left) and $Z_{\rm O}-M_{500}$ (right) relations at $M_{500}=10^{14}M_\odot$ for the four considered radial ranges. The dashed-lines represent the best-fit relations as described by Eq.~(\ref{eq3}) with the parameters reported in Table~\ref{tb4}. The error bars quote $1\sigma$ uncertainty of the best-fit normalisations at each considered redshifts.}
 \label{fig8}
\end{figure*}
  \begin{table*}
 \caption{\label{tb4}
  Best-fitting parameters of the relation in Eq.~(\ref{eq3}) for the normalisation evolution of the $Z_{\rm Fe}-M_{500}$ and $Z_{\rm O}-M_{500}$ relations in the AGN simulation.}
 \begin{center}
 \begin{tabular}{|ccccc|}
 \hline
& Fe & & O  \\
 \hline
Radial Range & $\log_{10}A_{\rm Fe}$ & $B_{\rm Fe}$ & $\log_{10}A_{\rm O}$& $B_{\rm O}$\\ 
\hline

 $[0.0-0.1]R_{500}$: & $-0.090\pm0.016$  & $-0.22\pm0.03$   &  $-0.175\pm0.003$  & $-0.05\pm0.01$ \\
 $[0.1-0.5]R_{500}$: &  $-0.404\pm0.012$ & $-0.18\pm0.02$  & $-0.528\pm0.008$  & $-0.08\pm0.02$ \\
 $[0.5-1.0]R_{500}$: & $-0.623\pm0.008$  & $-0.29\pm0.02$  &  $-0.762\pm0.006$ & $-0.29\pm0.01$ \\
 $[1.0-2.0]R_{500}$: &  $-0.765\pm0.007$  & $-0.06\pm0.02$  & $-0.911\pm0.007$ & $-0.05\pm0.01$ \\

  \end{tabular}
 \end{center}
 \end{table*}
In Fig.~\ref{fig6} we show the results on the iron (upper panel) and oxygen (lower panel) mass relations for both AGN and CSF simulations. As the mass-metallicity relations in the CSF run cannot be simply described by a single power law, we opt to represent the CSF results by showing the median relations of $Z_{\rm Fe}-M_{500}$ and $Z_{\rm O}-M_{500}$ in 10 logarithmic mass bins. To characterise the mass-metallicity relations of the AGN clusters, we fit the simulated data extracted in a given radial range and at a particular time, to a formula similar to Eq.~(\ref{eq4}):
\begin{equation}
Z_{X}=Z_{0,X}\times\bigg(\frac{M_{500}}{10^{14}M_\odot}\bigg)^{\beta_{X}}, \label{eq2}
\end{equation}
where $X$ stands for either iron or oxygen. The best-fitting parameters of the $Z_{\rm Fe}-M_{500}$ and $Z_{\rm O}-M_{500}$ relations as well as the Spearman's rank correlation coefficients are reported in Table~\ref{tb3}.  We include the results for all four apertures and for all times from $z=0$ to $z=1.5$. 

Confirming the results shown in Section 3, the CSF simulation shows a significantly steeper metallicity profile compared to the AGN. In the central region, the CSF clusters on average exhibit $Z_{\rm Fe}$ and $Z_{\rm O}$ which is $\sim1.5$ times higher than the AGN objects while in the outskirts it is lower by a factor $\sim2.3$. We notice that the CSF metallicity appears to drop in low-mass systems not only in the cluster core ($r<0.1R_{500}$), as seen in Section 3.1, but also at larger radii. We provide a more quantitative discussion on the comparison between AGN and CSF mass-metallicity relations later in Section 4.2.

 Unlike the CSF case, the mass-metallicity relation for AGN simulations is well described by a single power law at all the considered radius and redshift ranges. In general, the behaviour of the two mass-abundance relations, $Z_{\rm Fe}-M_{500}$ and $Z_{\rm O}-M_{500}$, is very similar and almost independent of the radial range and redshift at which they are computed. The Spearman's rank correlation coefficients (Table~\ref{tb3}) for both relations indicate an anti-correlation between the metal abundances and the mass in the innermost region ($r<0.5R_{500}$), especially strong at the lowest redshifts ($z \leq 0.5$) where the correlation coefficients are around $0.5-0.6$ with extremely low null-hypothesis probability --- below $<10^{-11}$. At higher redshifts, the anti-correlation between the inner quantities reduces to mild ($|r_s| \sim 0.3-0.4$) but still with high significance (null-hypothesis below $10^{-3}$). Considering, instead, the metallicity abundances in the cluster outskirts at low redshifts, i.e. $z \leq 0.5$, we verify that they do not correlate with the total mass ($|r_s|$ between 0.05  and $0.2$ with a high probability --- above 40 per cent --- that the correlation value is obtained by chance). However, the slopes of all relations are very shallow  ($\beta <-0.15$) everywhere and at anytime (see also Fig.~\ref{fig7}). The net difference between a $10^{14} M_{\odot}$ group and a $10^{15} M_{\odot}$ cluster is limited to an abundance decrease of 20-30 per cent. This mild trend is present at any time since the slope $\beta$  does not substantially vary with redshift as shown in the top panels of  Fig.~\ref{fig7}.


In the bottom panels of Fig.~\ref{fig7}, we report the evolution of the scatter. We notice that the $Z_{\rm O}-M_{500}$ relation always presents a higher intrinsic scatter than the $Z_{\rm Fe}-M_{500}$. The difference is stronger in the core ($r<0.1R_{500}$) where it reaches approximately $0.02-0.03$ dex. With the exception of the difference in amplitude, the trends with radii and time of both scatters are almost identical: the largest values are detected in the core and at most recent times ($\sigma_{\log_{10}Z|M}=0.14-0.16$ dex). Outside that region, both scatters promptly decrease and they remain constant out to the virial region ($\sim 2 R_{500}$). The redshift-dependence of the intrinsic scatters is present only in the core where the most important astrophysical processes take place. Specifically, for the iron and the oxygen abundances we respectively find a scatter increase of $37$ per cent and $18$ per cent from $z=1.5$ to $z=0$. 

\vspace{0.5 cm}

Since the slopes of the AGN mass-metallicity relations are constant with time at all radii with the exception of a minimal variation in the most external radial bin, we can study the evolution of the relations by simply measuring the shift in the normalisation at a fixed mass that we chose to be equal to $M_{500}=10^{14}M_\odot$. We quantify the evolution in each radial range by fitting the  best-fitting normalisations with the following relation:
\begin{equation}
Z_{0}(z)=A\times(1+z)^B\label{eq3}.
\end{equation}
The results are reported in Table~\ref{tb4} and shown  in Fig.~\ref{fig8}. The overall metallicity exhibits an extremely weak evolution within $R_{500}$ and no evolution at all in the region between $R_{500}$ and $2R_{500}$. The largest evolution regards the increase by about 23 per cent of the iron and oxygen abundance with respect to the present values in the $[0.5-1] R_{500}$ region. At redshift $1.5$ the normalisations of the iron and oxygen are already very close to the $z=0$ levels. To be precise, in the four radial ranges from the core to the virial radius, the iron normalisation at $z=1.5$ amounted to 82, 85, 77, 95 per cent of its value at $z=0$. The level of the oxygen is even less evolving since its normalisation is almost constant (in the four regions the $z=1.5$ value is 96, 93, 77, 96 per cent the $z=0$ value). Considering the shallow slope of the relation this result implies that the metal level was already built up in high$-z$ clusters.

\subsubsection{Evolution of the stellar fraction}
In the previous section, we found a mild variation in the oxygen abundance in the AGN simulation only in the $[0.5-1] R_{500}$ radial range. We check, here, whether this could be related to an increase of stellar content in that region. We, thus, consider how the stellar fraction, $f_{*}\equiv M_{\rm star}/M_{\rm tot}$, depends on the total mass, $M_{500}$, and how it evolves with time from $z=1.5$ to $z=0$. As found for the metal abundances and as expected, the stellar fraction exhibits negative gradient with radius: the stellar content is mostly concentrated in the central region ($r<0.1R_{500}$) and rapidly decreases towards the outskirts (see also, e.g., \citealt{planelles.etal.2013, battaglia.etal.2013}). In the core $r<R_{500}$ there is a trend with the cluster mass but no significant evolution from $z=1.5$ to $z=0$. Precisely, the stellar fraction is 10-15 per cent in the core of the smallest systems and about $4-5$ per cent in the largest clusters. The change of the relation normalisation, measured at $10^{14} M_{\odot}$, is less than 8 per cent. The result found in our simulations is in line with the recent observational study by \cite{chiu.etal.2018} (see also, e.g., \citealt{lin.etal.2012}), who also found an anti-correlation between stellar fraction and cluster mass within $R_{500}$ (see also, \citealt{planelles.etal.2013}) and no evidence of evolution in the redshift range between  $z=0.2$ and $z=1.25$.

\begin{figure}
\begin{center}
{\includegraphics[width=0.5\textwidth]{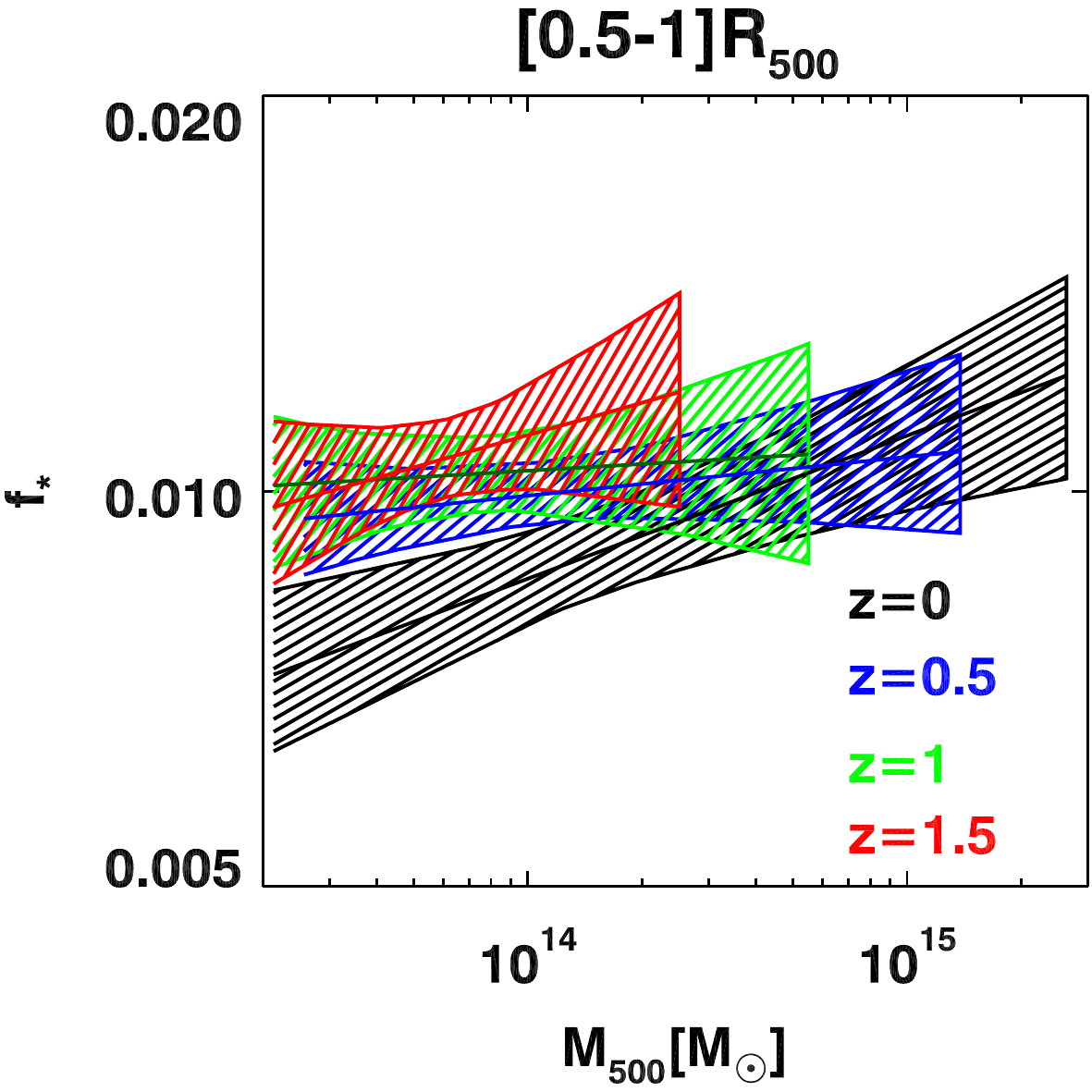}}
\end{center}
\caption{The stellar fraction-mass relation ($f_*-M_{500}$) in the AGN simulation shown at the radial range $[0.5-1]R_{500}$ for different redshifts. The solid lines represent the best-fit relations at those redshifts, while the shaded areas specify the $68.3\%$ confidence regions.}
 \label{fig9}
\end{figure}

However, if we restrict to the region $[0.5-1] R_{500}$ we do see that $10^{14} M_{\odot}$ clusters have a $30$ per cent reduction in their stellar fraction (see Fig.~\ref{fig9}).
The amplitude of this variation is still small but, noticeably, goes in the opposite direction with respect to the evolution of the oxygen abundance that we discuss in the previous section:
while, at fixed mass, the abundance grows from z=1.5 to z=0, the stellar content decreases. The increase in metallicity therefore cannot be associated to fresh stellar formation, instead, it is related to the accretion of already  
enriched gas, in addition to the accretion of pristine gas. In the next section, we investigate the role played by the AGN in raising the metal level in the outskirts of small groups. 
\begin{figure*}
\begin{center}
{\includegraphics[width=\textwidth]{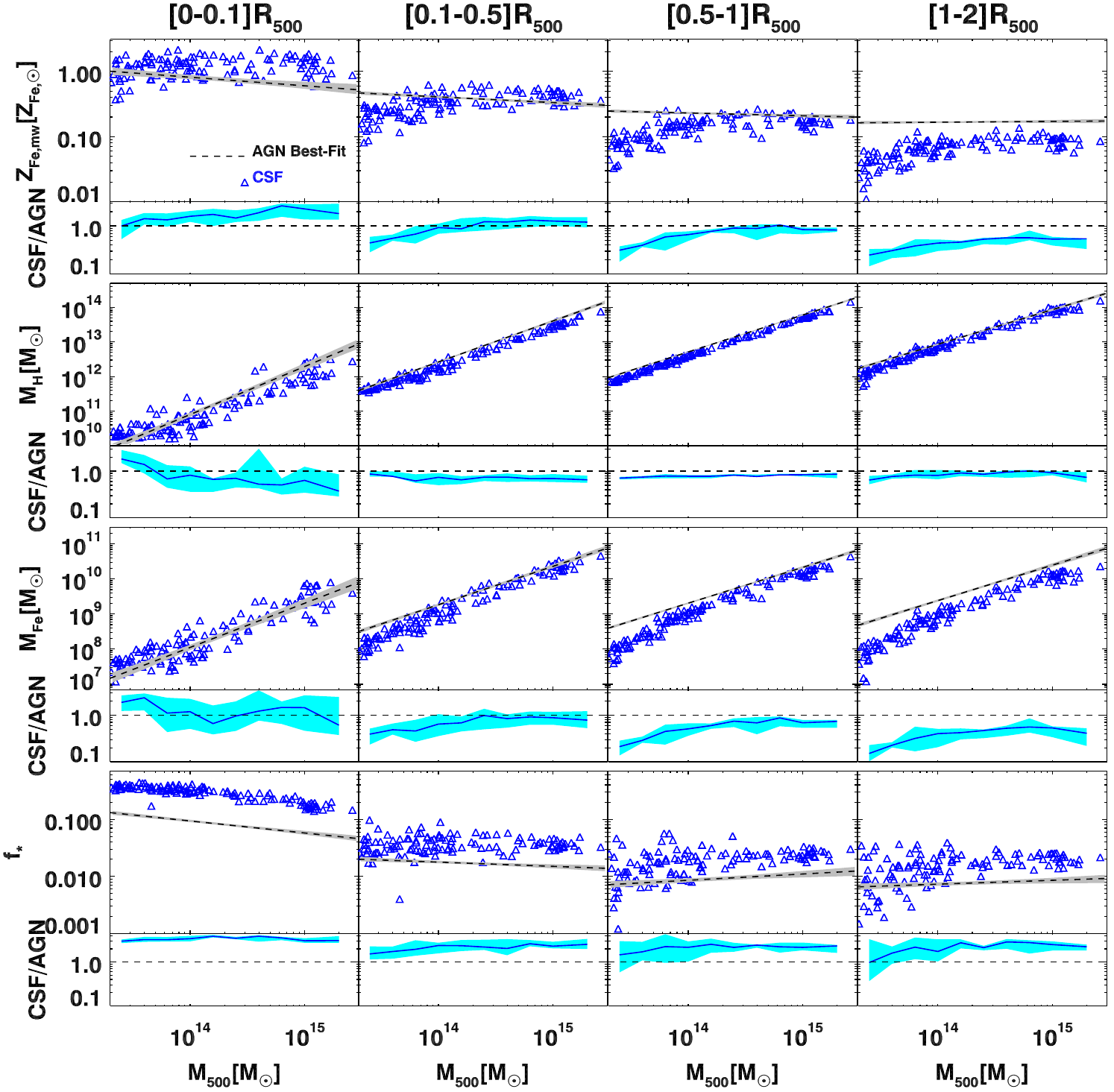}}
\end{center}
\caption{Comparison between AGN and CSF simulations at $z=0$ for the relation between cluster total mass ($M_{500}$) and various quantities: iron abundance ($Z_{\rm Fe}$), hydrogen mass ($M_{\rm H}$), iron mass ($M_{\rm Fe}$), and stellar mass fraction ($f_*$), for the four considered radial ranges. From top to bottom we show the $Z_{\rm Fe}-M_{500}$, $M_{\rm H}-M_{500}$, $M_{\rm Fe}-M_{500}$, and $f_{*}-M_{500}$ relations, respectively. In the top sub-panels, the AGN results are represented by best-fit relations (dashed line) and the grey regions specify the $68.3\%$ confidence regions. The bottom sub-panels show the median ratio (blue solid line) of the CSF value to the AGN best-fit relation as well as the $68.3\%$ confidence region (cyan).}
 \label{fig10}
\end{figure*}
\subsection{The effects of AGN feedback on the mass-metallicity relation}

Accounting that the iron and oxygen behaviours are very similar, in this Section we focus only on the former and we discuss the effect of the AGN on the mass-metallicity relation by comparing two sets of simulations obtained with and without the AGN.

In Fig.~\ref{fig10}, we show how the iron abundance, $Z_{\rm Fe}$, the hydrogen mass, $M_{\rm H}$, the iron mass, $M_{\rm Fe}$, and the stellar fraction vary with the total mass, $M_{500}$, in both simulations at $z=0$. The four columns correspond to the different radial apertures from the core to the virial regions (from left to right). In the bottom panels, we enlighten the differences between the CSF and AGN results. The most striking difference on the abundance-mass relations derived in the two runs is that without AGN the relation cannot be represented by a single power-law since there is a break at around $ 2\times 10^{14} M_{\odot}$.

At the cluster scale ($M_{500}>2\times 10^{14}M_\odot$), the abundance dependence on mass shows the same trend in the two runs at all radii, but there is a clear offset between the normalisations of the two relations. In the core, $r<0.1R_{500}$, the CSF $Z_{\rm Fe}$ is about 60 per cent higher than in the AGN run while in the two most external regions, $[0.5-1]R_{500}$ and $[1-2]R_{500}$, it is about 25 and 50 per cent lower, respectively.
 To understand the origin of these ratios, we look at the two separate contributions from the hydrogen mass and the iron mass in the second and third row, respectively. The gas fraction of the CSF clusters is always lower than that of the AGN clusters because the efficient radiative cooling, not regulated by the AGN activity, cools down a greater amount of hot gas that is subsequently converted into stars and thus is removed from the ICM. The phenomenon is in place at all radii but it is particularly strong in the core where the CSF runs host unrealistically massive bright central galaxies. 
 The result of the excessive overall star formation is to drastically enhance the stellar fraction, as can be seen in the fourth row. In the core and at fixed mass, the CSF runs can have three times more stars than in the AGN runs outside the core. In the CSF runs the star formation is actually still active at low-redshifts, however, the process does not lead to larger iron mass of the ICM  with respect to the AGN clusters (third row), because the freshly formed metals are immediately locked back into newly formed stars. In this way the efficient stellar production of the CSF runs prevents the circulation of the metals from the star forming regions to the ICM.
 For this reason the iron mass is lower than in the AGN outside the core. On the other hand, AGN feedback peaks at high redshift, when the potential well of the (proto-)cluster is still relatively shallow and star formation is also quite intense. As a consequence, this feedback channel plays a key role on spreading the metals created at high redshift. The process has the twofold effect of  removing metals from star forming regions and of enriching the pristine gas that surrounded the small potential well of the early galaxies and that subsequently accrete into the low-$z$ clusters \citep{biffi.etal.2017, biffi.etal.2018}.

Both radiative cooling and AGN heating have a stronger impact in the lowest mass regime. The iron abundance in the CSF groups is largely reduced everywhere outside the core, while it agrees with the value in the  AGN run in the innermost region. There, the hydrogen and iron masses depart from the AGN runs but with similar amplitude and sign. Indeed, generally the early activity of the AGN produces less concentrated groups with a reduced gas contribution. In addition to reducing the gas available for producing new stars, the AGNs also heat the medium. Both phenomena quench star formation, thereby reducing the accretion of already enriched gas, so that iron abundance can only grow through the explosions of long-lived SNIa. Outside the core, the reduction in the gas mass in the CSF groups is comparable to that in the CSF clusters. However, the iron mass decreases even further in groups without AGN because the radiative cooling, which is more efficient in smaller systems, selectively removes highly enriched gas.

\section{Discussion}
\subsection{Systematics}
 Cosmological simulations are useful to study the trend of metallicity with radius, mass, or temperature as well as its scatter and redshift evolution. The level of metals produced, instead, depends on the assumptions and simplifications underlying the stellar evolution and chemical models, and the measuring procedure. We dedicate this section to the discussion of systematics that might affect our conclusions.
 

{\it Model of chemical enrichment.} The chemical distribution and evolution derived from numerical simulations depend on the assumed models of star formation and stellar evolution. In \cite{tornatore.etal.2007}, the authors present the chemical model employed in our simulations and thoroughly investigate the effect of changing the stellar initial mass function, the yields for SN, the SN explosion rate (or equivalently the lifetimes), and the SN feedback efficiency. We briefly summarise here the main results of that work. Changing the IMF has the strongest effect on the pattern of chemical enrichment. Using an IMF that is top-heavier than Salpeter IMF produces a higher value of iron abundance (by a factor of 2), to a level that exceeds the observed amount, and also increases the $[{\rm O/Fe}]$ relative abundance ($\sim60\%$). On the other hand, the SN yields and explosion rate have marginal effect on the overall pattern of chemical enrichment. Finally, increasing the SN feedback strength results in suppressing the star formation rate thereby decreasing the level of metal enrichment. 


{\it Emission-weighted versus spectral metallicity.} To compare with the observational data, we consider the emission-weighted metallicity. In \cite{rasia.etal.2008} it was shown that the metallicity derived from {\it XMM-Newton} spectra of simulated mock observations was generally in good agreement with the emission-weighted metallicity. In particular, the emission-weighted estimation was proven to well reproduce the iron abundance for objects with temperature $T<2$ keV and $T>3$ keV where the Fe abundance is solidly measured via either Fe-L or Fe-K lines. A possible overestimate of order of 15-40 per cent was found in systems with intermediate temperature (with the highest discrepancy due to low signal-to-noise ratio). However, the small detected bias was found for the spectroscopic metallicity obtained under the assumption of a single temperature emitting plasma. Instead, the observational measures that we compare with (\citealt{mernier.etal.2018}) use a multi-temperature approach. With this approach, Mernier et al., appropriately capture the level of the continuum close to the line, therefore, suppress any residual bias in the equivalent width of the lines related to an incorrect determination of the continuum. Moreover, this approach allows one to reproduce the spectra of systems that behave as single-temperature objects, since the temperature parameters associated with their different model components are independent and free to converge if necessary (see Section 2.2 in \citealt{mernier.etal.2018}). 


{\it Projection effects.} Observational measurements of metallicity are rarely de-projected. Therefore, when we compare simulations with the observational data, we need to project the simulated quantities. The main effect of projecting the abundances along a certain direction is that the normalisation of the metallicity-temperature relation becomes slightly lower than in the case of 3D abundance. This effect is stronger in the cluster core ($r<0.1R_{500}$), while it becomes less important in the outer regions (see details in Appendix A1). Due to the negative radial gradient of metallicity (\citealt{biffi.etal.2017}), projection reduces the metal abundance with respect to the value computed within a sphere. This effect becomes more significant in low-mass objects as the majority of metals is concentrated in the very central regions. As a result, the 2D metallicity-temperature relation appears to be slightly shallower than the 3D relation, yet the two slopes are consistent within $1\sigma$. In conclusion, projection effects do not alter the trends discussed here.

\subsection{Comparison to other numerical studies}
Other authors recently also investigated the ICM metallicity-temperature relation using semi-analytical models (\citealt{yates.etal.2017}) as well as cosmological simulations (\citealt{dolag.etal.2017,barnes.etal.2017,vogelsberger.etal.2018}).

\cite{yates.etal.2017} compiled 10 different observational datasets taken from literature and homogenised the datasets to study the $Z_{\rm Fe}-T$ relation within $R_{500}$ and compare it with numerical results obtained from the semi-analytical L-GALAXIES galaxy evolution model. The semi-analytical model predicts a weak anti-correlation between the ICM iron abundance and temperature for groups and clusters as observed in our study (with the slope of the $Z_{500}-T_{500}$ relation of $\sim-0.1$), while this behaviour is present only in observed clusters of the homogenised dataset with $T>1.7$ keV. In the group regime, instead, the observed $Z_{\rm Fe}$ appears to drop as the temperature decreases. The authors suggest that the discrepancy between simulated and observed results can be solved by requiring an efficient mechanism to remove metal-rich gas, e.g., via AGN feedback, out of the central cluster regions. Our results do not support this interpretation. Indeed, our AGN simulations shows that the metallicity level in groups and clusters is similar in regions outside the cluster core ($r>0.1R_{500}$). Further on, the agreement between our AGN simulations and the CHEERS sample on the trend of the $Z_{\rm Fe}-T$ relation in the innermost regions suggests that the discrepancy between simulated and observed data reported in \cite{yates.etal.2017} might be due to a bias on the spectroscopic metallicity derived by fitting the observational spectra with an old atomic data code. \cite{mernier.etal.2018} discuss in detail how the old version of atomic data might significantly bias low the inferred metallicity in low-temperature systems. 

 	There are also studies of the $Z_{\rm Fe}-T$ relation using cosmological simulations: Magneticum (\citealt{dolag.etal.2017}), Cluster-EAGLE (\citealt{barnes.etal.2017}), and IllustrisTNG (\citealt{vogelsberger.etal.2018}). Those simulations include a wide range of astrophysical processes: radiative cooling, star formation and chemical enrichment, stellar and AGN feedback. \cite{barnes.etal.2017} and \cite{vogelsberger.etal.2018} predict relatively flat trend of the $Z_{\rm Fe,500}-T_{500}$ relation with $1\sigma$ dispersion less than $0.1$ solar unit. Similarly, \citealt{dolag.etal.2017} show very mild temperature trend of the Fe abundance in the central regions ($r<R_{2500}$). Those results are in line with our study on the trend of the $Z_{\rm Fe}-T$ relation.


The fact that our AGN simulations reproduce a trend for the iron abundance-temperature relation which is consistent with the latest observational results by \cite{mernier.etal.2018} supports current models of cosmological simulations, in particular it further emphasises the role of AGN feedback in shaping not only ICM X-ray properties but also its metal enrichment.

\section{Summary and Conclusions}

X-ray observations have shown that the intra-cluster gas metallicity appears to be distributed uniformly at large radii and exhibits no significant trend over cosmic time. 
We confirm these findings in our simulations (\citealt{biffi.etal.2017, biffi.etal.2018}) and we investigate, here, how the ICM metallicity depends on cluster scale 
(expressed both as temperature and mass) and how these relations evolve.
The simulations are performed with an updated version of the \texttt{GADGET-3} code which includes radiative cooling, star formation, metal enrichment, stellar and AGN feedback. Results from observational data are taken from \cite{mantz.etal.2017} and from the CHEERS sample (\citealt{de_plaa.etal.2017}). The latter is analysed with the last updated atomic model for the metallicity estimation (\citealt{mernier.etal.2018}). In the first part of our analysis, we compare our simulation results to the observational datasets and investigate how the ICM iron abundance varies as a function of the gas temperature. We particularly focus on potential biases, regarding the diversity of the core properties (CC and NCC)  and selection effects, that might affect the observational analyses. In the second part, we carry out a study of the mass-metallicity-redshift relation for iron and oxygen  abundances. In addition, we also investigate how the mass-metallicity relation behaves when we remove the AGN feedback, in order to investigate its effect on the ICM metal enrichment. In our discussion we highlight some of the model systematics that might affect the comparison between simulated and observed data and stress how the level of metallicity is sensitive to the numerical choices linked to stellar and chemical sub-grid models. For this reason, the analysis presented here is devoted to predicting and comparing general trends between global quantities involving metal abundance, rather than emphasising the absolute enrichment level. Throughout the study, we express both simulated and observed metallicity in terms of solar metallicity as obtained from \cite{asplund.etal.2009}. The main results of our study can be summarised as follows:
\begin{enumerate}
\item We compared the simulated $Z_{\rm Fe}-T$ relation of the AGN simulations to the CHEERS sample which spans from groups to hot clusters. Both datasets  show no evidence for a significant correlation between the ICM iron abundance and the ICM temperature measured in the core ($r<0.1R_{500}$) of systems with $T_{[0-0.1]R_{500}}>0.7$ keV. In particular, we did not find any significant break or feature that was present in earlier X-ray analysis. 
When we split our simulated sample in CC and NCC clusters, the former subsample better agrees with the CHEERS data. In particular both simulated and observed cluster cores consistently show a mean value of $Z_{\rm Fe}$ of about $0.75\ Z_{\rm Fe,\odot}$ with a dispersion of $40\%$ and $30\%$, respectively. We remind that even if the agreement between simulated and observed values of the iron abundance in the core might depend on the chemical model assumptions, the different metallicity levels between the simulated CC and NCC is a solid result and confirms the often claimed trend found in observational samples. Fitting the data with a power-law, we find a very shallow slope ($\sim-0.1$) implying an extremely small variation in the abundances of groups and clusters (20-30 per cent). 
\item When compared to the observed sample of massive clusters, with $T_{[0.1-0.5]R_{500}}>5$ keV and $0<z<1.2$, from \cite{mantz.etal.2017}, we find that the AGN simulations consistently reproduce the radial dependence of the $Z_{\rm Fe}-T$ relation for different radial ranges. For this sample of hot clusters, both simulated and observed data show a stronger correlation between iron abundance and temperature. However, we find that the correlation is significantly reduced when including smaller temperature systems. Our simulations reveal no significant trend with redshift (with a variation lower than $20\%$ since $z=1$) in the $Z_{\rm Fe}-T$ normalisation among all the considered radial ranges. This is at variance with respect to the results by \cite{mantz.etal.2017}, that present a stronger increase ($\sim40\%$) of iron abundance at intermediate radii, $0.1R_{500}<r<0.5R_{500}$. We notice, however, that among various observational works there is no agreement on this trend (see \citealt{ettori.etal.2015,McDonald.etal.2016}).
\item Both the iron and oxygen abundances of the AGN clusters exhibit an anti-correlation with cluster mass for regions within $R_{500}$, while no correlation is found in the cluster outskirts ($R_{500}<r<2R_{500}$). However, fitting the relation with a power-law, we found that even in the core the slope is shallow ($|\beta| < 0.15$) implying that the metallicity is only few tens of per cent different from poor groups to rich clusters.  We do not detect any significant evolution for the relation since  $z=1.5$. Considering that the metallicity varies little across the mass scale and does not change in time, we can conclude that the majority of the iron and oxygen were already reaching the current levels since high-$z$ throughout the clusters. 
\item The effect of the AGN feedback is studied by comparing the $Z_{\rm Fe}-M_{500}$ relation obtained in runs performed with and without the AGN feedback. 
Without the AGN, the $z=0$ systems present an higher stellar fraction at all radii but particularly in the core. The increase of the stellar fraction at each distance is almost independent on the mass of the system. The much higher stellar production, counter intuitively, is not associated to an increase of the iron mass of the ICM due to the fact that freshly produced metals are locked back into newly formed stars. Instead, the iron mass is comparable to the AGN case only in the core, but it is always reduced in the outskirts. This gap is scale-dependent being more extreme in the group regime, whereby causing the non-AGN iron abundance outside of the core to drop for systems with mass below $M_{500}\approx2\times10^{14}\ M_\odot$.

\end{enumerate}

Our study shows that simulations and observations agree in supporting a weak variation of the ICM metallicity from groups to clusters of galaxies and that the metal content does not substantially varies with time. In addition, we confirm that when AGN feedback is included the level of metallicity in the outskirts is flat and at a relatively high level (about 20-30 per cent of the solar value). All these findings further support the early enrichment scenario.
 
 There is still ample space for future works from both simulation and observation sides to consolidate the results. Future observations will be needed to improve the current statistics in the group regime. With the current X-ray telescopes, it is challenging to increase the number of observed small systems (e.g., those with temperature below 1 keV) as they are extremely faint in the X-ray band. In this regard, the next generation of X-ray telescopes, such as {\it ATHENA}\footnote{http://www.the-athena-x-ray-observatory.eu/}, would be of extremely utility in observing low-mass systems (\citealt{nandra.etal.2013, pointecouteau.etal.2013}). The {\it ATHENA} telescope, with large effective area and high-resolution spectroscopy, is expected to provide robust estimation of the gas-phase metallicity in the group regime.

From the numerical point of view, a direction of improvement would be represented by a more sophisticated metal diffusion model that now is implemented by spreading the metals to gas particles within the kernel. In addition, we have not included dust formation and destruction which should be considered for an accurate description of the metal content. Also, the current model of AGN feedback is exclusively thermal, while the mechanical AGN feedback (e.g., jets outflows) is relevant for anisotropically ejecting metals at larger radii from the core.



\section*{ACKNOWLEDGEMENTS}
We would like to thank the anonymous referee for constructive comments and suggestions that improve the paper. This work was supported by the Lend\"ulet LP2016-11 grant awarded by the Hungarian Academy of Sciences. We acknowledge financial support from the agreement ASI-INAF n.2017-14-H.0, the INFN INDARK grant, and �Consorzio per la Fisica� of Trieste. M.G. is supported by NASA through Einstein Postdoctoral Fellowship Award Number PF5-160137 issued by the Chandra X-ray Observatory Center, which is operated by the SAO for and on behalf of NASA under contract NAS8-03060. Support for this work was also provided by Chandra grant GO7-18121X. SP is  ``Juan de la Cierva'' fellow (ref. IJCI-2015-26656) of the {\it Spanish Ministerio de Econom{\'i}a y Competitividad} (MINECO) and acknowledges additional support from the MINECO through the grant AYA2016-77237-C3-3-P and the Generalitat Valenciana (grant GVACOMP2015-227). DF acknowledges financial support from the Slovenian Research Agency (research core funding No. P1-0188).

\bibliographystyle{mnbst}
\bibliography{ref}

\appendix
\section{Projection and Centring Effects}
In this section, we quantify how the metallicity-temperature relation can be affected by projection effects or by the adopted centre definition.

\subsection{Projection Effects}
\begin{figure*}
\begin{center}
{\includegraphics[width=\textwidth]{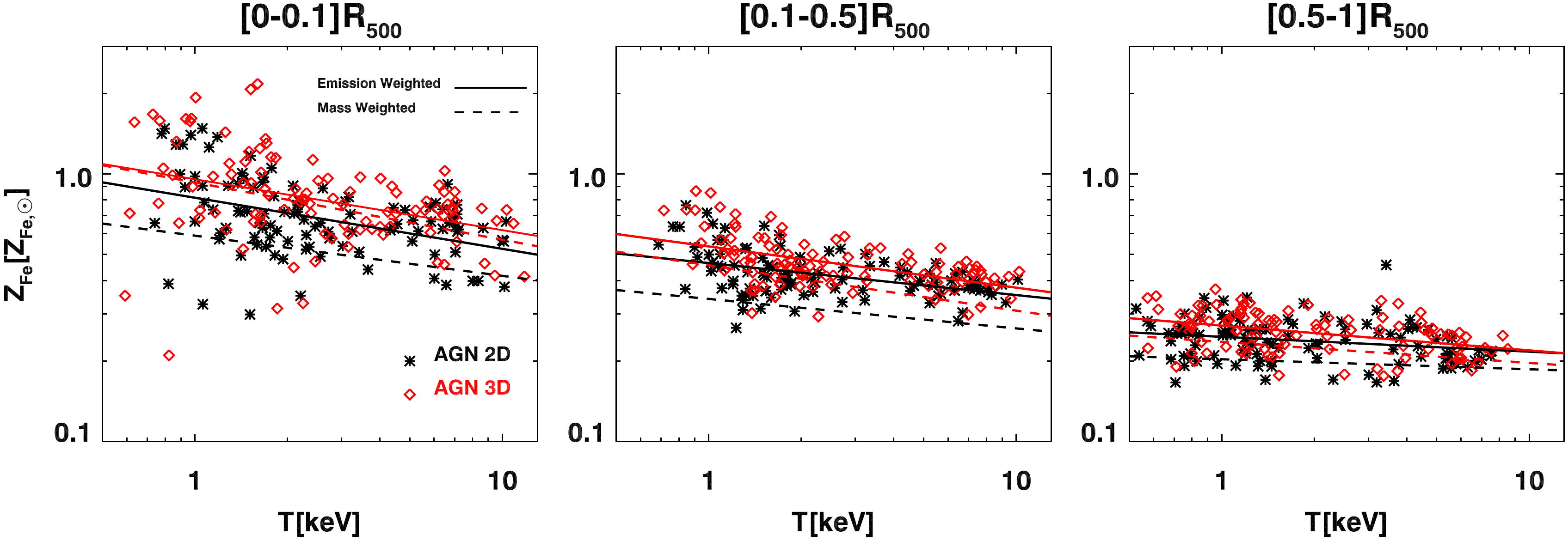}}
\end{center}
\caption{Comparison between 3D and 2D iron abundance as a function of spectroscopic-like temperature in the AGN simulation at $z=0$. The $Z_{\rm Fe}-T$ relation is shown for simulated clusters with a minimum number of 100 gas particles in the central region ($r<0.1R_{500}$) and with the mass $M_{500}>10^{13}\ M_\odot$. The individual data points represent the emission-weighted $Z_{\rm Fe}-T$ relations shown along with their best-fit relations (solid lines) obtained from fitting simulated data to Eq.~(\ref{eq4}), while for the sake of brevity the mass-weighted $Z_{\rm Fe}-T$ relations are represented only by their best-fit relations (dashed lines). All the best-fit parameters are reported in Tab. \ref{tba1}.}
 \label{figa1}
\end{figure*}

We show in Fig.~\ref{figa1} the comparison between the 3D and the 2D iron abundance as a function of the gas temperature in the AGN simulation at $z=0$. For this comparison, we employ  the spectroscopic-like estimate of the temperature, while for the iron abundance we use both mass-weighted and emission-weighted estimates, as described in Section 2.2. We fit simulated data, for those systems that have a minimum number of 100 gas particles in the central region ($r<0.1R_{500}$) and with the mass $M_{500}>10^{13}\ M_\odot$, to Eq.~(\ref{eq4}) to obtain the best-fit parameters for both $Z_{\rm Fe}-T$ relations. The best-fit parameters are reported in Table ~\ref{tba1}. We find that the emission-weighted $Z_{\rm Fe}-T$ normalisation is always higher than the mass-weighted one. The difference is higher in the 2D comparison reaching a level of $20-35$ per cent, while it is reduced for the 3D case to values of $15-20$ percent in the intermediate and outer regions, and to a minimum of 5 per cent in the innermost region of the core. The slopes are consistent at $1\sigma$. We limit the following discussion on the projection effects to the emission-weighted $Z_{\rm Fe}-T$ relations only.

Compared to the 3D $Z_{\rm Fe}-T$ relation, the 2D relation has slightly lower normalisation with a more marked difference in the central regions ($r<0.1R_{500}$) where the 3D normalisation is about $1.2$ times larger than the 2D value at the fixed temperature $T=1.7\ \rm{keV}$. While in the outer region ($0.5R_{500}<r<R_{500}$), the corresponding ratio of the 2D to the 3D values is $\sim1.07$. In particular, we notice that the offset between 2D and 3D iron abundance is more visible in low-mass systems than in more massive ones. As a consequence, the slope of the 2D $Z_{\rm Fe}-T$ relation is slightly shallower than the one of the 3D relation, yet the two slopes are consistent at $1\sigma$. We verify that the effect comes primarily from the discrepancy between 2D and 3D iron abundance with the 2D $Z_{\rm Fe}$ being lower than the 3D value due to the contribution of metal-poor gas particles along the projected direction. On the other hand, there is no significant difference between 2D and 3D spectroscopic-like temperatures. 

 \begin{table*}
 \caption{\label{tba1}
  Best-fit parameters of the relation in Eq.~(\ref{eq4}) for 2D and 3D $Z_{\rm Fe}-T$ relations in the AGN simulation at $z=0$ for simulated clusters with a minimum number of 100 gas particles in the central region ($r<0.1R_{500}$) and with the mass $M_{500}>10^{13}\ M_\odot$.}
 \begin{center}
 \begin{tabular}{|cccc|}
 \hline
Radial Range & $\log_{10}(Z_{0,{\rm Fe}}\ [Z_\odot])$ & $\beta_{\rm Fe}$ & $\sigma_{\log_{10}Z_{\rm Fe}|M}$ \\
\hline
{\bf 2D emission-weighted $Z_{\rm Fe}-T$}\\
$[0.0-0.1]R_{500}$: & $-0.132\pm0.015$ & $-0.19\pm0.04$ & $0.132\pm0.010$ \\
$[0.1-0.5]R_{500}$: & $-0.360\pm0.009$ & $-0.12\pm0.03$ & $0.076\pm0.006$ \\
$[0.5-1]R_{500}$: & $-0.620\pm0.008$ & $-0.06\pm0.02$ & $0.076\pm0.006$ \\
{\bf 2D mass-weighted $Z_{\rm Fe}-T$}\\
$[0.0-0.1]R_{500}$: & $-0.265\pm0.012$ & $-0.15\pm0.03$ & $0.104\pm0.008$ \\
$[0.1-0.5]R_{500}$: & $-0.492\pm0.007$ & $-0.11\pm0.02$ & $0.057\pm0.004$ \\
$[0.5-1]R_{500}$: & $-0.701\pm0.006$ & $-0.04\pm0.02$ & $0.056\pm0.004$ \\
{\bf 3D emission-weighted $Z_{\rm Fe}-T$}\\
$[0.0-0.1]R_{500}$: & $-0.069\pm0.019$ & $-0.19\pm0.05$ & $0.161\pm0.012$ \\
$[0.1-0.5]R_{500}$: & $-0.306\pm0.009$ & $-0.16\pm0.03$ & $0.081\pm0.006$ \\
$[0.5-1]R_{500}$: & $-0.589\pm0.007$ & $-0.09\pm0.02$ & $0.071\pm0.005$ \\
{\bf 3D mass-weighted $Z_{\rm Fe}-T$}\\
$[0.0-0.1]R_{500}$: & $-0.081\pm0.016$ & $-0.22\pm0.04$ & $0.140\pm0.011$ \\
$[0.1-0.5]R_{500}$: & $-0.381\pm0.008$ & $-0.17\pm0.02$ & $0.067\pm0.005$ \\
$[0.5-1]R_{500}$: & $-0.644\pm0.006$ & $-0.08\pm0.02$ & $0.061\pm0.005$ \\
  \end{tabular}
 \end{center}
 \end{table*}
  
\subsection{Miscentring Effects}
\begin{figure*}
\begin{center}
{\includegraphics[width=\textwidth]{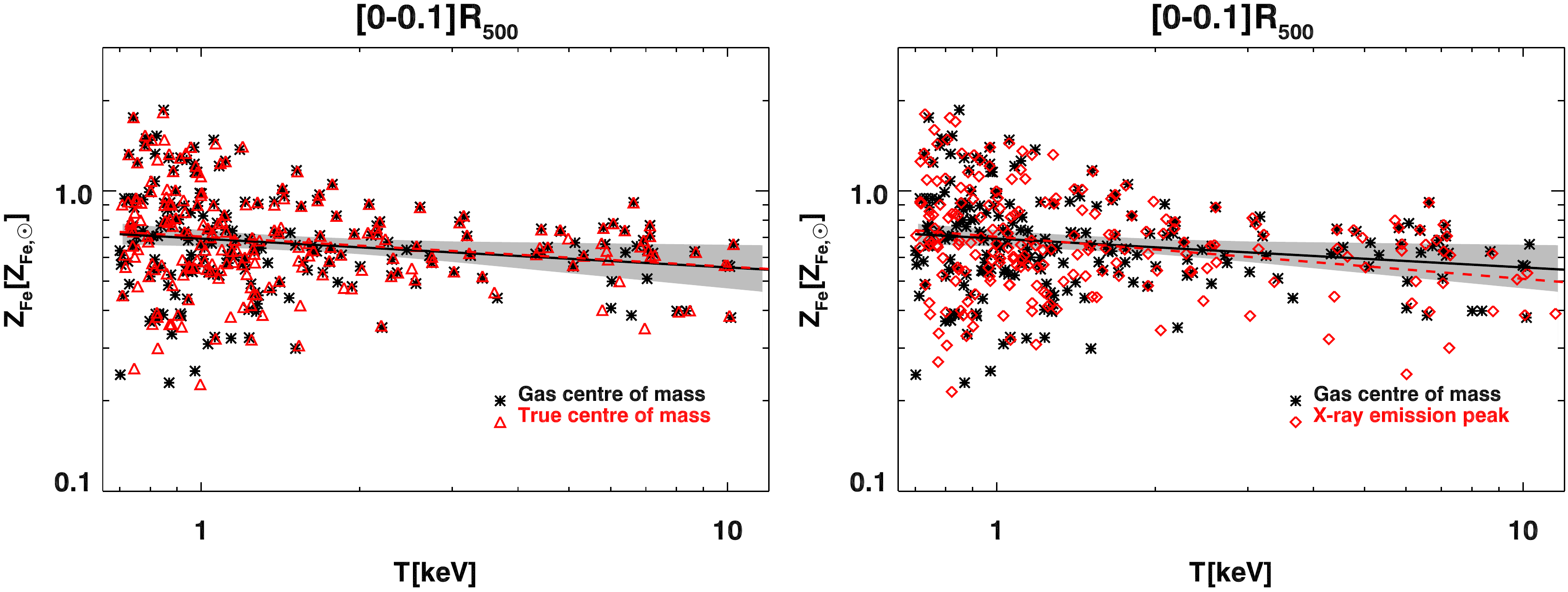}}
\end{center}
\caption{Comparison the 2D emission-weighted metallicity-temperature relation using the gas centre of mass to the relations using the true centre of mass (left) and the X-ray emission peak (right), in the AGN simulation at $z=0$. The relations are shown for the central region ($r<0.1R_{500}$). The solid line and the grey-shaded area representing the best-fit relation and the $68.3\%$ confidence level, respectively, are identical to the ones shown in Fig. \ref{fig2} (for the full AGN sample). The red dashed lines are the best-fit relations with parameters reported in Tab. \ref{tba2}.}
 \label{figa2}
\end{figure*}

We show in Fig. \ref{figa2} the comparison among the metallicity-temperature relations obtained by employing three different centre definitions: the minimum of the cluster potential well, the centre of the gas mass within $R_{2500}$, and the X-ray emission maximum within the same radius.  For this comparison, we used the 2D emission-weighted $Z_{\rm Fe}-T$ relations in the AGN simulation at $z=0$. Fig. \ref{fig2} and the best-fit parameters reported in Table \ref{tb1} (the full AGN sample) refer to the centre defined as the centre of the gas mass. We also characterise the two other $Z_{\rm Fe}-T$ relations by fitting them to Eq. (\ref{eq4}) and report the best-fit parameters in Table {\ref{tba2}}. As shown in Fig. \ref{figa2}, the three metallicity-temperature relations are in total agreement among each other. From the tables, we can notice that normalisation, slope and scatter are consistent within the 1$\sigma$ uncertainties. We conclude that the results on the metallicity-temperature relation shown in the paper are robust against the centre definition used.

\begin{table*}
 \caption{\label{tba2}
  Best-fit parameters of the relation in Eq.~(\ref{eq4}) for 2D $Z_{\rm Fe}-T$ relations in the AGN simulation at $z=0$ centring at the true centre of mass and the X-ray emission peak, as shown in Fig. \ref{figa2}.}
 \begin{center}
 \begin{tabular}{|cccc|}
 \hline
Centre & $\log_{10}(Z_{\rm 0,Fe}\ [Z_\odot])$ & $\beta_{\rm Fe}$ & $\sigma_{\log_{10}Z_{\rm Fe}|M}$ \\
\hline
{\bf True centre of mass} & $-0.175\pm0.011$ & $-0.10\pm0.03$ & $0.155\pm0.008$ \\
{\bf X-ray emission peak} & $-0.185\pm0.012$ & $-0.14\pm0.04$ & $0.167\pm0.008$ \\
  \end{tabular}
 \end{center}
 \end{table*}

\end{document}